\documentclass[reqno,keywordsasfootnote]{article}
\usepackage[english]{babel}
\usepackage[latin1]{inputenc}
\usepackage{amsmath,amsthm,amsfonts,amssymb,times}
\numberwithin{equation}{section}
\usepackage[final]{epsfig}
\usepackage{color}

  \usepackage{epstopdf}
\newtheorem{theorem}{Theorem}[section]
\newtheorem{proposition}[theorem]{Proposition}
\newtheorem{lemma}[theorem]{Lemma}

\theoremstyle{definition}
\newtheorem{definition}{Definition}[section]

\newcommand{\Z}{ {\mathbb Z} }
\newcommand{\N} {{\mathbb N}}

\newcommand{\R}{\mathbb{R}}

\let\Im\undefined
\let\Re\undefined
\DeclareMathOperator{\Im}{Im}
\DeclareMathOperator{\Re}{Re}

\makeatletter
\let\Im\undefined
\let\Re\undefined
\DeclareMathOperator{\Im}{Im }
\DeclareMathOperator{\Re}{Re }
\DeclareMathOperator{\indfct}{\rm 1}
\DeclareMathOperator{\dist}{\rm dist}
\DeclareMathOperator{\Tr}{\rm Tr}
\def\Plim{\mathop{\rm \mathbb{P}\!-\!\lim\; }}

\makeatother
\def\be{\begin{equation} }
\def\ee{\end{equation} }
\begin{document}
\date{\small \today}

\title{The Canopy Graph and Level Statistics for Random Operators on Trees}

\author{Michael Aizenman and Simone Warzel\\ \small Departments of
Mathematics and Physics,\\ \small Princeton University, Princeton NJ
08544, USA.}%


\maketitle

\begin{abstract}
For operators with homogeneous disorder, it is generally expected
that there is a relation between the spectral characteristics of a
random operator in the infinite setup and the distribution of the
energy gaps in its finite volume versions, in corresponding energy
ranges. Whereas pure point spectrum of the infinite operator goes
along with Poisson level statistics, it is expected that purely
absolutely continuous spectrum would be associated with gap
distributions resembling the corresponding random matrix ensemble.
We prove that on regular rooted trees, which exhibit both spectral 
types, the eigenstate point process has always Poissonian limit.
However, we also find that this does not contradict the picture
described above if that is carefully interpreted, as the relevant
limit of finite trees is not the infinite homogenous tree graph but rather
a single-ended  ``canopy graph''.  For this tree graph, the
random Schr\"odinger operator  is proven here to have only
pure-point spectrum at any strength of the disorder.    For more 
general single-ended trees  it is shown that the spectrum is always 
singular -- pure point possibly with singular continuous component 
which is proven to occur in some cases.
\\[2ex]

(2000 Mathematics Subject Classifiction: 47B80, 60K40)\\
\vspace*{1cm}
\end{abstract}


\newpage

\tableofcontents

\newpage

\section{Introduction}

\subsection{An overview}

For random operators with extensive disorder it is generally
expected that there is an interesting link between the nature of the
spectra of the infinite operator and the statistics of energy gaps of
the finite-volume restriction of the random operator.  Extensively
studied examples of operators with disorder include the
Schr\"odinger operator with random potential \cite{CaLa90,PF,Stoll}
and the quantum graph operators, as
in~\cite{KS99,ASW05b}.
The often heard conjecture (see eg.\ \cite{AS86,Sh93,Ef97,DR03} and
references therein) is that on the scale of typical
energy spacing the energy levels will exhibit Poisson statistics
throughout  the pure point (pp) spectral regimes,  and level repulsion
through energy ranges for which the infinite systems has absolutely
continuous (ac) spectrum.

The presence of pp spectra for random Schr\"odinger operators
on $\ell^2(\Z^d)$ or $L^2(\R^d)$  is now thoroughly investigated.
In this context, the conjectured
Poisson statistics has been established throughout the localization
regime for the lattice cases \cite{Min96}, and also for the $d=1$
continuum operators \cite{Mol81}, which exhibit only pure point
spectra.
For dimensions $d>2$ it is expected that random operators will exhibit
also ac spectra.  However, so far the only cases
of operators with extensive disorder for which the existence of
an {\em ac} spectral component was proven are operators on tree
graphs~\cite{Kle95,Kle98,ASW05,FrHaSp06}.
Attempting to analyze the conjecture in that context we encountered
two surprises, on which we would like to report in this note:
\begin{enumerate}
\item For random operators on trees, under  an auxiliary technical 
assumption which is spelled below,
the level distribution is given by
Poisson statistics through the entire spectral regime.   In
particular, the statistics of the neighboring levels is typically free of level
repulsion even throughout the spectral regimes  where the infinite
tree operator has {\em ac} spectrum.
\item  For the purpose of the level statistics of finite tree graph
operators, as observed within energy windows scaled by a volume
factor,  the relevant infinite graph is not the regular tree graph,
but another one, which is called here the {\em canopy graph}.
This graph is isomorphic to the horoball subgraphs of the regular tree,
in the terminology explained in \cite{Woe00}.
\end{enumerate}

The second point is related to the known result, proven in
\cite[Thm.1.1]{Sz89,Sz90}, concerning the density
of states of Schr\"odinger operators on hyperbolic spaces (see 
Section~\ref{sect:dos}).

The main surprise (1.) is then somewhat reconciled with the above general
expectation by the next result:
\begin{enumerate}
\item[3.] The corresponding random operator on the (infinite) canopy
graph, has only pp spectrum at any non-zero level of
extensive disorder.
\end{enumerate}

In the above statement, the absence of an absolutely continuous component is
readily explained by the fact that the canopy graph has exactly one end -
in the sense (see, e.\ g.\ \cite{Woe00})  that from each point on it 
emanates exactly one infinite path.
However, some more detailed analysis is required to prove that the 
spectrum is pure point.  To make that clear we also prove the 
following, which may be
of some independent interest:
\begin{enumerate}
\item[4.] There are tree graphs with exactly one end on which the spectrum of
the corresponding Schr\"odinger operator almost surely has a singular 
continuous
spectral component.
\end{enumerate}
 
We shall now make those statements more explicit.

\subsection{The random operator on finite subgraphs of a regular tree}

Let $\mathcal{T} $ denote the vertex set of a rooted tree graph  for
which all vertices have $ K $ neighbors in directions away from the
root~$ 0 $, for some fixed  $ K \geq 2 $.
Out of the infinite tree $ \mathcal{T} $ we carve an increasing
sequence of finite trees of depth $L$, denoting:
\be
        \mathcal{T}_L := \left\{ x \in \mathcal{T} \, : \, \dist(0,x)
\le  L \right\} \, ,
\ee
Here  $ \dist(\cdot,\cdot) $ refers to the natural distance between
two vertices in $ \mathcal{T} $.
The adjacency operator on the Hilbert space of square-summable
functions $\ \psi \in \ell^2(\mathcal{T}_L)$ is given by
\begin{equation}\label{eq:adjac}
     \left(A  \psi\right)(x) := \mkern-10mu \sum_{\substack{y \in
\mathcal{T}_L \, : \\ \dist(x,y) = 1} }\mkern-10mu \psi(y)  \, .
\end{equation}
In the notation for $A$
we omit the index ($\mathcal{T}_L$) indicating on what $\ell^2$-space the
operator acts.
We will be concerned with random perturbations of the adjacency
operator, namely self-adjoint operators of the form  \\ [1 ex]
\begin{equation} \label{defH}
     H_{\mathcal{T}_L} := A  +  V  + B   \,
\end{equation}
acting in
$ \ell^2(\mathcal{T}_L) $, with
$ V $ a random potential and $B$ a boundary term, both given
by multiplication operators:
\begin{align}
    \left(V \psi\right)(x) &:= \omega_x \psi(x) \, , \\[1ex]
    \left(B \psi\right)(x) & :=  \left\{ \begin{array}{cl} b \,  \psi(x) &
      \mbox{if $\dist(0,x) =  L $} \\
     0  & \mbox{otherwise} \, \qquad \, .
         \end{array}\right.     \label{eq:bc}
    \end{align}
     Here $ \{ \omega_x \}_{x \in \mathcal{T}} $
stands for a collection of independent identically distributed (iid)
random variables, and  $ b \in \mathbb{R} $ is a fixed number.  The
latter serves as a control parameter, in effect allowing to vary  the
boundary conditions at the {\em outer boundary}, a term by which we
refer to the set  $\partial \mathcal{T}_L := \{ x \in \mathcal{T} \,
: \,  \dist(0,x) =  L \}$.\\

Throughout this discussion we restrict ourselves to random
potentials whose probability distribution meets the following condition:
\begin{enumerate}
\item[]{\bf Assumption A1:}~~ The distribution of the potential variables
$\omega_x$  is of bounded density, $ \varrho \in L^\infty(\mathbb{R})
$, and satisfies
     $ \int_\mathbb{R} | \omega_0 |^\tau \varrho(\omega_0) d
\omega_0 < \infty $ for
some $ \tau > 0 $.
\end{enumerate}

The main object of interest will be the random point process of
eigenvalues of $ H_{\mathcal{T}_L} $, seen on the scale of the mean
level spacing.    For a finite operator, the expected number of
eigenvalues  in an interval is proportional to the number of sites of
the finite graph, $ | \mathcal{T}_L | $ (see the Wegner estimate
\eqref{eq:Wegner} below).
It is therefore natural to consider the point process of the
eigenvalues as seen under the magnification by the volume.
Thus, for a given energy $ E \in \mathbb{R} $ we consider the random
point measure
\begin{equation}\label{def:mu}
     \mu_{L}^{E} := \sum_n \delta_{\, |\mathcal{T}_L|
\left(E_n(\mathcal{T}_L) - E \right)} \, ,
\end{equation}
where $ \{ E_n(\mathcal{T}_L) \} $ denotes the sequence of
random eigenvalues of $ H_{\mathcal{T}_L} $, counting multiplicity.\\

Our main results are derived under the additional assumption:
\begin{enumerate}
\item[] {\bf Assumption A2:}~~ The expectation values $ \mathbb{E}\left[\ln
     \left| \big\langle \delta_0 , \big( H_{\mathcal{T}_L} - E
\big)^{-1} \delta_0 \big\rangle\right|\right]  $ are
     equi\-continuous functions of $ E \in I $ over some Borel set $ I 
\subset \mathbb{R} $.
\end{enumerate}
An explicit example, which satisfies both Assumptions~{\bf A1}
and {\bf A2} for $ I = \R $,
is the Cauchy distribution.  In that case, Cauchy integration
allows to equate the above expectation value with the resolvent
at an energy off the real axis, and then {\bf A2} is easily seen to be valid.
For the general case, through a Thouless-type formula one gets
     \be
      \mathbb{E}\left[\ln
     \left| \big\langle \delta_0 , \big( H_{\mathcal{T}_L} - z
\big)^{-1} \delta_0 \big\rangle\right|\right]
         = \Re \int_\mathbb{R} \frac{\nu_L(E)}{E-z} \, dE
     \ee
     where $\nu_L(E) := \mathbb{E}\left[\Tr
P_{(-\infty,E)}(H_{\mathcal{T}_L}) \right]
     - K \, \mathbb{E}\left[\Tr
P_{(-\infty,E)}(H_{\mathcal{T}_{L-1}}) \right] $ defines the spectral
shift function related to the removal of the root in $ \mathcal{T}_L $.
The symbol $ \Tr $ refers to the trace and $ P_I $ denotes the 
spectral projection
onto the Borel set $ I \subset \mathbb{R} $.
Assumption~{\bf A2} is therefore
connected to the regularity of this spectral shift function.
     Such regularity may be deduced from some of the results in \cite{AK92}
which address distributions ``near'' the Cauchy case.\\

The main result of the present paper is
\begin{theorem}[Poisson statistics]\label{thm:main}
Let $H$ be a  random Schr\"odinger operator $H$, as in  \eqref{defH}, 
for which the conditions  {\bf A1} and {\bf A2 } hold  for some 
interval $ I \subset \mathbb{R}$. Then for Lebesgue almost
every $ E \in I $  the random point measures
     $     \mu_{L}^{E}   $ converges
     to a Poisson point measure $ \mu^E $ as $ L \to \infty $.
\end{theorem}

The intensity of the limiting Poisson point process $ \mu^E $ is given by the
Lebesgue measure times the
canopy density of states  $ d_\mathcal{C}(E) $, which is the
the topic of Subsection~\ref{subsec:intensity}.
In particular, it is shown there that this intensity is non-zero in
some energy regimes.
The proof of Theorem~\ref{thm:main} is provided in 
Section~\ref{sec:Proofmain} below.
The convergence refers to the usual notion
of weak convergence of random point measures \cite{Kal02}.\\

As explained above, at first glance Theorem~\ref{thm:main} may
appear to be surprising, since it is known that  random
Schr\"odinger operators on regular infinite trees exhibit also
spectral regimes where the spectrum is ac~\cite{Kle95,Kle98,ASW05,FrHaSp06}.
Furthermore, the cases for
which this result was established include some for which both
assumptions are satisfied, and
the ac spectrum was even shown to be pure in the
present setting \cite{Kle98}.  Thus, the result may appear to fly in
the face of the oft repeated expectation that ac spectra of the
infinite volume limit should be linked with level repulsion of the
finite subsystems.
However, that discrepancy is resolved by the observations presented next.

\subsection{The canopy operator} \label{sect:dos}
\begin{figure}[hbt]
\begin{minipage}{0.5\textwidth}
\begin{picture}(0,0)%
\includegraphics{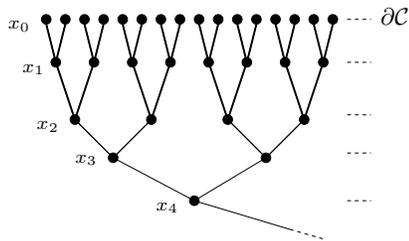}%
\end{picture}%
\setlength{\unitlength}{1579sp}%
\begingroup\makeatletter\ifx\SetFigFont\undefined%
\gdef\SetFigFont#1#2#3#4#5{%
    \reset@font\fontsize{#1}{#2pt}%
    \fontfamily{#3}\fontseries{#4}\fontshape{#5}%
    \selectfont}%
\fi\endgroup%
\begin{picture}(5850,3685)(-224,-3083)
\put(826,-1861){\makebox(0,0)[lb]{\smash{{\SetFigFont{7}{8.4}{\rmdefault}{\mddefault}{\updefault}{\color[rgb]{0,0,0}$x_3$}%
}}}}
\put(-224,239){\makebox(0,0)[lb]{\smash{{\SetFigFont{7}{8.4}{\rmdefault}{\mddefault}{\updefault}{\color[rgb]{0,0,0}$x_0$}%
}}}}
\put(
1,-436){\makebox(0,0)[lb]{\smash{{\SetFigFont{7}{8.4}{\rmdefault}{\mddefault}{\updefault}{\color[rgb]{0,0,0}$x_1$}%
}}}}
\put(226,-1336){\makebox(0,0)[lb]{\smash{{\SetFigFont{7}{8.4}{\rmdefault}{\mddefault}{\updefault}{\color[rgb]{0,0,0}$x_2$}%
}}}}
\put(2101,-2611){\makebox(0,0)[lb]{\smash{{\SetFigFont{7}{8.4}{\rmdefault}{\mddefault}{\updefault}{\color[rgb]{0,0,0}$x_4$}%
}}}}
\put(5626,314){\makebox(0,0)[lb]{\smash{{\SetFigFont{9}{10.8}{\rmdefault}{\mddefault}{\updefault}{\color[rgb]{0,0,0}$\partial
\mathcal{C} $}%
}}}}
\end{picture}%
\end{minipage}
\begin{minipage}{0.5\textwidth}
\caption{The canopy graph $
\mathcal{C} $ for $ K = 2 $.
The dots indicate that the boundary layer $ \partial \mathcal{C} $ as
well as any layer below is infinite.
The vertices $ x_0, x_1, x_2, \dots $ mark the points on the unique
path $ \mathcal{P}(x_0) $ of $ x_0 $ to
``infinity''.}\label{Fig:Canopy}
\end{minipage}
\end{figure}

It may seem natural to take the line that the infinite-volume limit
of the sequence of finite trees $\mathcal{T}_L$ is the
infinite tree $\mathcal{T}$.  That is indeed what the graph converges
to when viewed from the perspective of the root, or from any site at
fixed distance from the root.  However, if one fixes the perspective
to be that of a site at the outer boundary of  $\mathcal{T}_L$, the
limit which emerges is  different.
We use the term canopy tree to
describe that limiting graph.
More explicitly, the rooted canopy tree $ \mathcal{C} $
is recursively defined in terms of a hierarchy of infinite layers of vertices:
starting from an infinite outermost boundary layer,
$ \partial \mathcal{C} $, each layer is partitioned into sets of $K$ elements,
and the elements of each component are joined to a common site in the 
next layer;
see Figure~\ref{Fig:Canopy}.

Two remarks apply:
\begin{enumerate}
\item
The canopy graph can be imbedded in the regular tree.
It is isomorphic to a horoball, the canopy's outermost boundary layer
corresponding to a horosphere and its different layers to horocycles 
- in the terminology explained, e.g,, in \cite{Woe00}.
	\item
	The observation that a given nested sequence of graphs may have
different limits applies also to other graphs.  In particular,
for the sequence $[-L,L]^d \cap \Z^d$ analogs of the canopy
construction yield the  graphs $\N\times \Z^{(d-1)}$, and also
$\N^k \times \Z^{(d-k)}$ for any $0 \le k \le d$.
\end{enumerate}

In view of the multiplicity of the limiting graphs, one may ponder
which of them is of relevance for a given question.
If the question concerns an extensive quantity, e.g.
$\Tr F(H_L) = \sum_x \langle \delta_x\, , \, F(H_L) \,\delta_x\rangle $,
where $H_L$ is a finite-range
operator and $F$ some smooth function, then the answer depends on
  how the environment appears from the perspective of a point which
is chosen at random uniformly within the finite graph.
In this respect there is a fundamental difference between the finite
cubic subgraphs of $\Z^d$, $[-L,...,L]^d $, and the finite subgraphs
of a regular tree.
In the former case, for $L\to \infty$, under the uniform sampling
the distance  from the boundary regresses to infinity,  and $\Z^d$ is
the natural limit.   However, for the tree graphs  $\mathcal{T}_L$
the distribution of the distance to the boundary converges to the
exponential distribution:  the fraction of points whose distance to
the outer boundary exceeds $n$, decays as $K^{-n}$.   In this case it
is the canopy graph which captures the limit.
For an explicit formulation of the statement we introduce the 
\emph{canopy operator} acting on
$ \ell^2(\mathcal{C}) $,
\begin{equation}
     H_\mathcal{C} := A  + V + B \, ,
\end{equation}
Here,  $ A $ is the adjacency operator on $ \ell^2(\mathcal{C}) $
which is defined similarly to \eqref{eq:adjac}, and $B$ is a
boundary term acting as in \eqref{eq:bc}, with
the same $ b \in \mathbb{R} $.
Moreover, the iid random variables $ \{ \omega_x \}_{x\in
\mathcal{C}} $ underlying the random multiplication operator $ V $ are
supposed to satisfy {\bf A1}.

Associated to $ H_\mathcal{C} $ is the following \emph{density of
states (dos) measure} given by
\be \label{dos}
     n_\mathcal{C}(I) := \frac{K-1}{K}
     \sum_{n=0}^\infty K^{-n} \, \mathbb{E}\left[ \langle
\delta_{x_n} \, , P_{I}(H_{\mathcal{C}}) \,  \delta_{x_n} \rangle
\right] \, ,
\ee
where the sum ranges over all vertices $ x_0 , x_1 , \dots $ on the
unique path $ \mathcal{P}(x_0) $ of a given vertex $ x_0 \in \partial
\mathcal{C} $ to infinity,
see Figure~\ref{Fig:Canopy}.
Note that $ n_\mathcal{C} $ does depend on the choice of the boundary 
conditions
parameter $ b \in
\mathbb{R}$, however it is independent of the choice of $ x_0 \in \partial
\mathcal{C} $ on the boundary.

\begin{theorem} \label{thm:DOS}
The density of states of the finite tree ${\mathcal T}_L$ is asymptotically
given by $ n_\mathcal{C} $  in the sense that, with probability one:
for any  bounded continuous  $ F \in C_b(\mathbb{R}) $
\be  \label{eq:dos}
\lim_{L \to \infty} |\mathcal{T}_L|^{-1} \Tr F( H_{\mathcal{T}_L}) \
= \ \int_\mathbb{R} n_\mathcal{C}(dE) \, F(E) \, .
\ee
\end{theorem}
The statement reflects the fact that on trees, asymptotically, almost all points 
are
located not far from the surface. The proof is given in 
Subsection~\ref{App:Erg2}.
The fact that for non-amenable graphs like trees
	bulk averages as in \eqref{eq:dos}  do not converge to corresponding
	infinite-volume quantities is well known.
In particular,  a continuum analogue of Theorem~\ref{thm:DOS} was presented in
\cite[Thm.1.1]{Sz89,Sz90} where it is shown that the finite-volume 
density of states  (dos) of a Laplacian plus Poissonian random 
potential on hyperbolic space converges to
the dos on a horoball.   Analogous statements apply to
the dos of periodic Schr\"odinger operators
	on hyperbolic spaces \cite{AS93}. \\

Part of the suprise of Theorem~\ref{thm:main} is now removed by the
following result, which is proven in Section~\ref{subsec:Loccan}.
\begin{theorem}[Localization of canopy states]\label{thm:loc}
If the conditions {\bf A1} and {\bf A2 } hold for $ I \subset \mathbb{R} $,
then the random canopy operator $ H_{\mathcal{C}} $ has almost surely
only pure point spectrum in $ I $.
\end{theorem}

It may also be of interest to note the following curious property of
${\mathcal C}$.

\begin{theorem}[Spectrum of the adjacency operator]\label{thm:adjopball}
The spectrum of of the adjacency operator with (constant)
     boundary conditions, $  A + B $ on $ \ell^2(\mathcal{C}) $, is
only pure point with compactly supported eigenfunctions.
\end{theorem}
A more detailed description of the spectrum of  the adjacency
operator can be found in Subsection~\ref{App:Erg}.
The considerations yielding Theorem~\ref{thm:adjopball} are
essentially as in the analysis of the regular tree in \cite{AF00}.

We postpone further comments on possible directions for studies of 
the relation which was explored in this work to  the concluding 
section, Sect.~\ref{sec:comments}.  

\section{Conditions for Poisson statistics for tree operators}

\subsection{The density bounds of Wegner and Minami}

Key information on the point process which describes the
eigenvalues of the random operator as seen under the magnification by 
the volume factor
$|\mathcal{T}_L|$ is provided in the following two essential
estimates.  The Wegner estimate implies that the mean density
of states is bounded relative to the Lebesgue measure.
The Minami bound
guarantees that the energy levels are non-degenerate on the scale
of the mean level spacing.

\begin{proposition} \label{prop:W+M}
Under the assumption~{\bf A1}, for every bounded Borel set $ I
\subset \mathbb{R} $ and every $ L \in \mathbb{N} $
     \begin{equation}\label{eq:Wegner}
     \mathbb{P}\left( \Tr  P_{ I}(H_{\mathcal{T}_L}) \geq 1
\right)  \leq \mathbb{E}\left[ \Tr  P_{ I}(H_{\mathcal{T}_L})\right]
      \leq | I | \, | \mathcal{T}_L | \, \| \varrho \|_\infty
\end{equation}
(the Wegner estimate), and
\begin{align}
\sum_{m = 2}^\infty \mathbb{P}\left( \Tr P_I(H_{\mathcal{T}_L}) \geq
m \right) & \leq \mathbb{E}\left[ \Tr P_I(H_{\mathcal{T}_L}) \, \big(
\Tr P_I(H_{\mathcal{T}_L}) - 1 \big)\right] \notag \\
     & \leq \pi^2 \, | I |^2 | \mathcal{T}_L |^2 \, \| \varrho
\|_\infty^2 \,  \label{prop:minami}
\end{align}
(the Minami estimate).
  Here $ \Tr P_I(H_{\mathcal{T}_L}) $ stands for the trace of the
spectral projection of $ H_{\mathcal{T}_L} $ onto $ I $.
\end{proposition}
A proof of the Wegner bound \eqref{eq:Wegner} can be found in \cite{Weg81,PF}.
Minami's estimate \eqref{prop:minami} is presented in \cite[Lemma~2, 
Eq.~(2.48)]{Min96}, see also \cite{GV06}.
Although it is stated there for $\Z^d$ only,
its derivation clearly applies to all graphs.

\subsection{Proof strategy and a sufficient condition}  \label{sec:sufficient}

The main effort in the proof of the convergence of the energy level
process to a Poisson limit is to establish infinite divisibility.
In the background are the following observations.
\begin{enumerate}
\item
The Wegner  estimate~\eqref{eq:Wegner} implies
tightness of the collection of random variables $ \{ \mu_L^E(I) \} $ 
for all intervals $ I \subset \mathbb{R} $ and
every $ E \in \mathbb{R} $.
This in turn guarantees that for every $ E \in \mathbb{R} $ the sequence of
measures $ \{ \mu_L^E \} $,
is tight with respect to the vague topology on
the space of Borel measures on the real line.
Since the subspace of point measures is closed with respect to this
topology, all accumulation points of the above sequence are point
measures \cite{Kal02}.
\item
In order to show that any accumulation point is a Poisson measure,
it is sufficient to prove that each such point is infinitely
divisible  and almost surely has no double points.  The latter is
guarantied already by the Minami estimate.
\item
Once the divisibility property is established, for convergence
  of the point process it suffices to show that the intensity measure
  of any accumulation point is given by some common measure  (\cite{Kal02}).
  In our case, that measure is the canopy mean density of states 
$n_{\mathcal C}$.
\end{enumerate}

Turning to the divisibility, one may note that for random operators 
on $\ell^2(\Z^d)$ the divisibility and convergence of the energy level
process to a Poisson process were proven by Minami under a natural 
localization condition (the fractional moment
characterization of the pp spectral regime \cite{Min96}).
However, Minami's proof does not extend to tree graphs, since it 
makes use of the fact that cubic regions in 
$ \Z^d $ have the van Hove property, 
which is that  most of the volume 
is, asymptotically, far from the surface.  While this approach 
does not apply to  trees, or hyperbolic 
spaces, with positive Cheeger isoperimetric constant, 
for trees there is another pathway towards infinite 
divisibility of any accumulation point of $ \{ \mu^E_L \}$.
In order to show $ K^N $-divisibility at some arbitrary $ N \in \mathbb{N} $,
we cut the finite tree $ \mathcal{T}_L $  below the $ N $th
generation. This leaves us with a ``tree stump'' and the subtrees
$\mathcal{T}_L(x) $ which are forward
to vertices $ x $ in the $ N $th generation.
Associated with the above collection of forward subtrees is the
collection of iid point measures
     \begin{equation}\label{def:appmu}
          \mu_{x,L}^{E} := \sum_n \delta_{\, |\mathcal{T}_L|
\left(E_n(\mathcal{T}_L(x)) - E \right)} \, .
     \end{equation}
For the sum $  \sum_{\dist(0,x)=N}\mu_{x,L}^{E}$ to be asymptotically 
equal to $ \mu^E_L $ as $ L \to \infty $, so that any of its 
accumulation points is
$ K^N $-divisible, it suffices that the spectral measures associated 
with the roots of the subtrees satisfy the following fluctuation 
condition.

For any site $ x \in \mathcal{T}_L $ the spectral measure is defined 
for Borel sets $ I \subset \R $ by
\begin{equation}
         \sigma_{x,L}(I) := \big\langle \delta_x,
P_I(H_{\mathcal{T}_L}) \,\delta_x \big\rangle \, .
\end{equation}
By a Wegner-type estimate the averaged spectral measure,
$ \mathbb{E}\left[ \sigma_{x,L} \right] $, is seen to be ac
with a density bounded uniformly in $ L \in \mathbb{N} $.
The condition we require for the proof of divisibility is that the 
typical value of $ \sigma_{x,L} $ on the
scale of its mean, $ |\mathcal{T}_L|^{-1} $, is much smaller than the 
average value.   More explicitly:

\begin{definition} 
\label{def:unif}
For a fixed site $ x \in \mathcal{T} $ and  energy $ E \in \mathbb{R} $,  
 the sequence of spectral measures $ 
\{ \sigma_{x,L} \} $ is said to   be
   \emph{negligible in probability}   
iff  for all $ w > 0 $
\begin{equation}\label{eq:condition}
     \Plim_{L\to \infty} | \mathcal{T}_L|\;\,  \sigma_{x,L}\big(E+ | 
\mathcal{T}_L|^{-1}\, (-w,w)\big) = 0 \, ,
\end{equation}
where the limit refers to distributional convergence.
\end{definition}
Several remarks apply:
\begin{enumerate}
\item
The prelimit quantity in \eqref{eq:condition} compares the spectral measure
$ \sigma_{x,L} $ to the blown-up Lebesgue measure of the 
corresponding interval.  In terms of the normalized eigenfunctions  $ 
\psi_n(\mathcal{T}_L) \in \ell^2(\mathcal{T}_L) $ of $ 
H_{\mathcal{T}_L} $, with the corresponding
eigenvalues $ E_n(\mathcal{T}_L) $,  one has:
\begin{equation}\label{eq:specmeas}
     \sigma_{x,L}(I)
     = \sum_{ E_n(\mathcal{T}_L) \in I} \left| \big\langle
     \delta_x, \psi_{n}(\mathcal{T}_L) \big\rangle \right|^2   \, .
\end{equation}
By Wegner's estimate \eqref{eq:Wegner} the mean number of levels 
within any given Borel set with Lebesgue measure proportional to
$ | \mathcal{T}_L |^{-1} $ is bounded uniformly in $L$.  If the 
corresponding eigenfuctions are spread uniformly over the volume, and 
the relevant spectral density is non-zero, then the above condition 
is not satisfied, since whenever an eigenvalue falls within the 
interval the rescaled spectral measure is not smaller than order $1$. 
Thus, condition ~\eqref{eq:condition} is equivalent to the statement 
that either {\em i.} the probability of finding an eigenvalue in the energy 
window vanishes, or {\em ii.} the corresponding  eigenfunctions are spread very 
unevenly over the volume, so  that typically
$ \left| \big\langle
     \delta_x, \psi_{n}(\mathcal{T}_L) \big\rangle \right|^2 \times 
|\mathcal{T}_L| \  << \ 1 $ even though the average of the quantity 
over the volume (or over $n$) is $1$.   In other words, on the scale 
of their mean the eigenfunctions 
exhibit \emph {divergent fluctuations}.
  \item
We shall show below, in Appendix~\ref{app:sing}, that the above 
scenario {\em i.} occurs if for some energy range 
the (weak) limiting measure, $ \big\langle \delta_x,
P_{\cdot}(H_{\mathcal{T}}) \,\delta_x \big\rangle = \lim_{L \to
\infty} \sigma_{x,L} $ for $ x \in \mathcal{T} $, is purely singular.  
In that case, due to the mutual singularity, 
the spectral measures underperforms at Lebesgue - chosen 
$ E \in \R $, in small intervals of arbitrary scale.  It is 
more of an issue to verify that \eqref{eq:condition} holds 
throughout  the  regime of ac spectrum of $ H_\mathcal{T} $.
This will be proven in Subsection~\ref{Subsec:deloc} below, where 
we show that for canopy graph scenario {\em ii.} is in effect.
\end{enumerate}

The criterion for Poisson statistics may now be formulated as follows.

\begin{theorem}[Condition for Poisson statistics]\label{prop:infdiv}
Suppose that the sequence of spectral measures at the root, $ \{ 
\sigma_{0,L} \} $,   is  {\emph negligible in probability}  at $ E 
\in \mathbb{R} $.
Then for any $ N \in
\mathbb{N} $ the sum $  \sum_{\dist(0,x)=N}\mu_{x,L}^{E}$ converges
     weakly to the same limit as $ \mu_{L}^E $, i.e., for all $ \psi 
\in  L^1_+(\mathbb{R}) $
\begin{equation}\label{eq:Ndiv}
     \lim_{L \to \infty} \left| \mathbb{E}\left[ e^{-
\sum_{\dist(0,x)=N}\mu_{x,L}^{E}(\psi)} \right]
     -  \mathbb{E}\left[e^{- \mu_{L}^{E}(\psi)} \right] \right| = 0 \, .
\end{equation}
As a consequence, all accumulation points of $ \mu_{L}^E $ are random 
Poisson measures.
\end{theorem}
\begin{proof}
Since the set of functions $ \varphi_z := \Im (\,\cdot-z)^{-1} $ with
$ z \in \mathbb{C}^+ $ are
dense in $ L^1_+(\mathbb{R})  $, it suffices to verify
\eqref{eq:Ndiv} for such functions.
It is easy to see that the latter follows from the distributional convergence
\begin{equation}\label{eq:distrconv}
     \Plim_{L \to \infty} \left| \; \sum_{\dist(0,x)=N}
\mu_{x,L}^{E}(\varphi_z) - \mu_{L}^E(\varphi_z) \right| = 0  \, .
\end{equation}
Abbreviating $ \xi_L := E +  z \, |\mathcal{T}_L|^{-1} $
the prelimit in \eqref{eq:distrconv} can be written as
\begin{multline}
    \frac{1}{|\mathcal{T}_L|} \left| \; \sum_{\dist(0,x)=N}  \Im \Tr \big(
H_{\mathcal{T}_L(x)} - \xi_L \big)^{-1}
     - \Im \Tr \big( H_{\mathcal{T}_L} - \xi_L \big)^{-1}  \right|\\
    \leq \frac{1}{|\mathcal{T}_L|} \sum_{|y| < N } \Im \big\langle
\delta_y ,  \big( H_{\mathcal{T}_L} - \xi_L \big)^{-1}
     \delta_y\big\rangle \\
    + \frac{1}{|\mathcal{T}_L|}\sum_{\dist(0,x)=N}  \, \Big|
     \sum_{ y \in \mathcal{T}_L(x)}   \big\langle \delta_y , \big(
H_{\mathcal{T}_L(x)} - \xi_L \big)^{-1}     \delta_y\big\rangle -
     \big\langle \delta_y ,  \big( H_{\mathcal{T}_L} - \xi_L
\big)^{-1}    \delta_y\big\rangle \Big| \, . \label{eq:split}
\end{multline}
The first term on the right side converges to zero in distribution as
$ L \to \infty $.
Using the resolvent identity twice the modulus in the second term is
seen to be equal to
\begin{multline}\label{eq:est}
         \Big| \sum_{ y\in \mathcal{T}_L(x)}
       \big\langle \delta_y , \big( H_{\mathcal{T}_L(x)} - \xi_L
\big)^{-1}     \delta_{x}\big\rangle \,
     \big\langle     \delta_{x^-} \big( H_{\mathcal{T}_L} - \xi_L
\big)^{-1}    \delta_{x^-} \big\rangle  \big\langle \delta_{x} ,
     \big( H_{\mathcal{T}_L(x)} - \xi_L \big)^{-1} \delta_y
\big\rangle \Big| \\
     \leq  \big\langle \delta_{x} , \big| H_{\mathcal{T}_L(x)} -
\xi_L \big|^{-2}     \delta_{x}\big\rangle
     \; \left| \big\langle     \delta_{x^-} \big( H_{\mathcal{T}_L}
- \xi_L \big)^{-1}    \delta_{x^-} \big\rangle \right|
      \, ,
\end{multline}
where $ x^- $ is the backward neighbor of $ x $ in $ \mathcal{T}_L $.
The second term in \eqref{eq:est} is bounded in probability as $ L
\to \infty $ as is seen from the fractional moment bound 
\eqref{eq:smombound} below.
Thanks the fluctuation condition and Lemma~\ref{lemma:equiv}
below the first term,
when dividing by $ | \mathcal{T}_L | $, converges to zero in this
limit.
\end{proof}
The previous proof was based on the following
\begin{lemma}\label{lemma:equiv}
Either of the following is equivalent to the statement that the 
sequence of spectral measures $ \{ \sigma_{x,L} \} $ is negligible 
in probability, at $ E \in \mathbb{R} $:
\begin{enumerate}
\item  For all $ \alpha \in \mathbb{R} $: $\quad\displaystyle
\Plim_{L\to \infty}  |\mathcal{T}_L|^{-1} \,
     \big\langle \delta_x , \big(H_{\mathcal{T}_{L}}- E -\alpha \,
|\mathcal{T}_L|^{-1}\big)^{-2} \delta_x \big\rangle = 0 $.
\item For all $ z \in \mathbb{C}^+ $:
     $\quad\displaystyle \Plim_{L\to \infty} \Im \big\langle
\delta_x , \big(H_{\mathcal{T}_{L}}- E - z \,
|\mathcal{T}_L|^{-1}\big)^{-1} \delta_x \big\rangle = 0 $.
\end{enumerate}
\end{lemma}
\begin{proof}
{\it 1. $\Rightarrow$ 2. $\Rightarrow $ 'negligibility in 
probability':} These two implications
are a consequence of the following chain of elementary inequalities
     \begin{align}
      2 \, (\Im z)^{-1} \, \big\langle \delta_x ,  P_{I(\Re z,\Im
z)}\big(H_{\mathcal{T}_L}\big) , \delta_x \big\rangle
     & \leq \Im \big\langle \delta_x ,  \big( H_{\mathcal{T}_L} -
z \big)^{-1} \delta_x\big\rangle  \notag \\
     & \leq \Im z  \, \big\langle \delta_x ,  \big(
H_{\mathcal{T}_L} - \Re z \big)^{-2} \delta_x\big\rangle ,
\label{eq:1folgt3}
     \end{align}
     valid for all $ z \in \mathbb{C}^+ $, where $ I(E,w) := E +
(-w,w) $ denotes the open interval centred at $ E $ of width $ 2 w >
0 $.\\[0.5ex]
{\it 'Negligibility in probability' $\Rightarrow$ 1.:} We split the 
prelimit in {\it 1.} into two
terms by inserting a spectral projection onto the interval
     $ I_L := I(E,w |\mathcal{T}_L|^{-1}) $ and its complement.
Abbreviating $ \xi_L := E + \alpha \, |\mathcal{T}_L|^{-1} $
     the first term is then estimated as follows
     \begin{equation}\label{eq:3folgt1}
      |\mathcal{T}_L|^{-1} \, \big\langle  \delta_x ,  \big(
H_{\mathcal{T}_L} - \xi_L\big)^{-2}  P_{I_L}(H_{\mathcal{T}_L}) \,
\delta_x\big\rangle
     \leq   |\mathcal{T}_L|^{-1}
     \frac{\big\langle \delta_x ,  P_{I_L}(H_{\mathcal{T}_L})
\delta_x\big\rangle}{\dist\left(\sigma(H_{\mathcal{T}_L}),\xi_L
\right)^{2}}\, .
     \end{equation}
     Using Wegner's estimate \eqref{eq:Wegner} and \eqref{eq:condition},
     this term is seen to converge in distribution to zero  as $ L
\to \infty $
     for any $ w > 0 $.
     The remaining second term is
     \begin{multline}
     |\mathcal{T}_L|^{-1}\, \big\langle  \delta_x ,  \big(
H_{\mathcal{T}_L} - \xi_L\big)^{-2}  P_{I_L^c}(H_{\mathcal{T}_L}) \,
\delta_x\big\rangle \\
     \leq 2 w^{-1} \, \Im \big\langle \delta_x ,
\big(H_{\mathcal{T}_{L}}-  \xi_L + i w \, |\mathcal{T}_L|^{-1}
\big)^{-1} \delta_x \big\rangle \, . \label{eq:remainder}
     \end{multline}
     The imaginary part of the resolvent is bounded in
probability.  Therefore the probability that the right side in
\eqref{eq:remainder}
     is greater than any arbitrarily small constant is arbitrarily
small for $ w $ large enough.
\end{proof}

\section{Decay estimates of the Green function}
\label{sec:decay}

As was shown in \cite[Thm.~II.1]{AM}, fractional moments of the
Green function of rather general random operators are uniformly
bounded.
\begin{proposition}[Fractional moment bounds] Under the
assumption~{\bf A1}, for any $ s \in (0,1) $,
\begin{equation}\label{eq:smombound}
    C_s := \sup_{z \in \mathbb{C}} \, \sup_{L \in \mathbb{N}} \, \sup_{
x, y \in \mathcal{T}_L} \mathbb{E} \left[ \left| \big\langle
\delta_x , \big( H_{\mathcal{T}_L} - z \big)^{-1} \delta_y
\big\rangle\right|^s  \big| \{x,y\}^c \right] < \infty \, ,
\end{equation}
where the average   $\mathbb{E}\left[ \,  \cdot  \, \big| \{x,y\}^c 
\right]   $ is the conditional expectation with respect to the 
sigma-algebra the generated by
$ \{ \omega_v \}_{v \in  \mathcal{T}\backslash\{x,y\}}$.
\end{proposition}

The main aim of this section is to prove that fractional moments of
the Green function of
$ H_{\mathcal{T}_L} $ are not only bounded but decay exponentially
along any ray in the tree.
\begin{theorem}[Exponential decay I] \label{lemma:decGreen}
Assume {\bf A1} and {\bf A2 } holds for a bounded Borel set $ I 
\subset \mathbb{R} $.
Then there exists $ s
\in (0,1) $, $\delta, C \in (0, \infty) $ such that for all $ E \in I
$, $ L \in \mathbb{N} $
and all $ x \in \mathcal{T}_L $ which are in the future of $ y \in
\mathcal{T}_L $
\begin{equation}\label{eq:decGreen}
     \mathbb{E}\left[ \left| \big\langle \delta_y , \big(
H_{\mathcal{T}_L} - E \big)^{-1} \delta_x \big\rangle\right|^s\right]
     \leq C \exp\left[ - s \big( \delta +  \ln \sqrt{K} \big) \,
\dist(x,y) \right] \, .
\end{equation}

\end{theorem}
Several remarks apply:
\begin{enumerate}
     \item Unlike on $\Z^d$, for trees the exponential
     decay~\eqref{eq:decGreen} does not
imply complete localization, i.e.\ dense pure
     point spectrum at all energies.  In fact,  the infinite-volume operator
     $ H_\mathcal{T} $ has a regime with delocalized
     eigenstates~\cite{Kle95,Kle98,ASW05,FrHaSp06}.
     \item The rate of decay in \eqref{eq:decGreen} is related to a
     Lyapunov exponent of the infinite-volume operator $
H_{\mathcal{T}} $, cf.\ Subsection~\ref{subsec:Lya} below.
     Note that in the unperturbed case where $ H_{\mathcal{T}} = A
$, the decay rate in \eqref{eq:decGreen} would be
given by $ \ln \sqrt{K} $. It is important for us that the decay rate in
\eqref{eq:decGreen} is strictly larger.
\end{enumerate}

\subsection{The decay of fractional moments}
Our proof of Theorem~\ref{lemma:decGreen} is based on  
reasonings similar to that which applies in 
in the one-dimensional setup,  
the reason being that any pair of sites on the tree is connected 
through a single path.  
As in one dimension, the Green function decays exponentially at a rate 
characterized by a Lyapunov exponent.
In order to relate the decay rate of    
the fractional-moment of the Green function to that exponent,
the following simple lemma will be of help.
\begin{lemma}\label{lemma:triv}
     Let $ (\xi_j)_{j=1}^N $ be a collection of independent,
positive random variables with
     $ c:= \max_{j} \mathbb{E}\left[ \left(\ln \xi_j\right)^2
\left( \xi_j + 1 \right)\right] / 2 < \infty $. Then $ X :=
\prod_{j=1}^N \xi_j $ satisfies
     \begin{equation}
         \mathbb{E}\left[ X \right] \leq \exp\left(
\mathbb{E}\left[\ln X \right]  + N c \,   \right) \, .
     \end{equation}
\end{lemma}
\begin{proof}
     This is a straightforward consequence of the assumed
independence and the elementary inequalities
     $ e^\alpha \leq 1 + \alpha + \alpha^2 \left( e^\alpha + 1
\right) / 2 $ and $ 1+ \beta \leq e^\beta $ valid for all $ \alpha$,
$ \beta \in \mathbb{R} $.
\end{proof}

\begin{lemma} \label{lemma:decGreen1}
Let $ I \subset \mathbb{R} $ be a bounded Borel set and assume {\bf 
A1}. Then for every $
\varepsilon > 0 $ there exists $ s_\varepsilon \in (0,1/2) $ and $
L_\varepsilon \in \mathbb{N} $
such that for all $ s \in (0, s_\varepsilon) $, $ E \in I $, $ L \geq
L_\varepsilon $ and $ x \in \mathcal{T}_L $ with $ \dist(0,x) \geq
L_\varepsilon $
\begin{multline}\label{eq:decGreen1}
     \ln \mathbb{E}\left[ \left| \big\langle \delta_0 , \big(
H_{\mathcal{T}_L} - E \big)^{-1} \delta_x \big\rangle\right|^s\right]
\\
     \leq \varepsilon \; \dist(0,x) + s \; \mathbb{E}\left[ \ln
\left| \big\langle \delta_0 , \big( H_{\mathcal{T}_L} - E \big)^{-1}
\delta_x \big\rangle \right|
          \right]  \, .
\end{multline}
\end{lemma}
The proof  is based on the factorization
of the Green function on a tree, which we recall from
\cite[Eq.~(2.8)]{Kle98},
\begin{align}\label{eq:Gfactor}
     & \big\langle \delta_0 , \big( H_{\mathcal{T}_L} - E
\big)^{-1} \delta_x \big\rangle = \prod_{j=0}^{\dist(0,x)}
\Gamma_{j,L} \notag \\
     & \qquad \mbox{with} \quad \Gamma_{j,L} := \big\langle
\delta_{x_j} , \big( H_{\mathcal{T}_L(x_j)} - E \big)^{-1}
\delta_{x_j} \big\rangle \, .
\end{align}
Here $0 =: x_0, x_1 , \dots , x_{ \dist(0,x) } := x $ are the
vertices on the unique path connecting the root $ 0 $ with $ x $.
Moreover,
$ \mathcal{T}_L(x_j) $ is that subtree of $ \mathcal{T}_L  $ which is
rooted at and forward to $ x_j $.

\begin{proof}[Proof of Lemma~\ref{lemma:decGreen1}]
The idea is to group together subproducts of \eqref{eq:Gfactor} und
use certain independence properties in order to apply
Lemma~\ref{lemma:triv}.
To do so we pick $ L_0 \in \mathbb{N} \setminus \{1\} $ and express
the distance of $ x $ to the root modulo $ L_0 $,
\begin{equation}
     \dist(0, x) = N_x L_0  + L_x
\end{equation}
with suitable $ N_x \in \mathbb{N}_0 $ and $  L_x \in \{0, \dots , L_0 -1 \} $.
Thanks to the factorization \eqref{eq:Gfactor} we may thus write
\begin{align}\label{eq:moreprod}
     & \left| \big\langle \delta_0 , \big( H_{\mathcal{T}_L} - E
\big)^{-1} \delta_x \big\rangle\right|^s = \left( \prod_{k=
0}^{N_x-1} X_k Y_k \right) \, R \\
     & \quad \mbox{with} \qquad X_k Y_k   := \prod_{j= k L_0
}^{(k+1) L_0 -1}\left| \Gamma_{j,L} \right|^s \, ,
         \quad  R := \prod_{j=N_x L_0}^{\dist(0,|x|)} \left|
\Gamma_{j,L} \right|^s \, .
\end{align}
Each product $ X_k Y_k $ may now be split into two terms by setting $
Y_k $ equal to the modulus of a diagonal element of the operator
corresponding to the forward subtree
$ \mathcal{T}_L(x_{k L_0}) $,
\begin{equation}
      Y_k :=  \left|\big\langle \delta_{x_{(k+1) L_0 -1}} \, ,
\big( H_{\mathcal{T}_L(x_{k L_0})} - E \big)^{-1} \delta_{x_{(k+1)
L_0 -1}} \big\rangle\right|^s \, .
\end{equation}
The point is that in this way we obtain a collection $ (X_k)_{k=0}^{N_x-1} $ of
independent, positive random variables. Moreover,
\begin{enumerate}
     \item each random variable $ X_k $ is independent of the
value of the potential at vertex $ x_j $ with $ j = (k+1) L_0 -1 $
for some
     $ k\in \{ 0, \dots , N_x-1\} $ or $ N_x L_0 \leq j \leq \dist(0,|x|) $.
     \item the random variable $  Y_k $ is independent of the
value of the potential at vertex $ x_j $ with $ 0 \leq j < k L_0 $.
\end{enumerate}
One may therefore succesively integrate the product in
\eqref{eq:moreprod} by first conditioning on the potential at $ 
x_{L_0 -1} $ thereby
integrating $ Y_0 $, then conditioning
on $ x_{2L_0 -1} $ thereby integrating $ Y_1 $ and so forth until we
reach $ x_{N_x L_0 -1} $ and integrate $ Y_{N_x-1} $. Thanks to
\eqref{eq:smombound}
these integrals are all uniformly bounded,
\begin{equation}
      \mathbb{E}_{x_{(k+1) L_0 -1}} \left[ Y_k \right] \leq C_s \, .
\end{equation}
Moreover, conditioning on the values of the potential at $ x_{N_x L_0} $
and $ x $, the fractional-moment bound \eqref{eq:smombound} also yields
\begin{equation}
       \mathbb{E}_{x_{N_x L_0},x}\left[R\right] \leq C_s \, .
\end{equation}
One is then left with the integral of the product $ \prod_{k=
0}^{N_x-1} X_k $, which can be bounded with the help of
Lemma~\ref{lemma:triv}.
Its assumption is satisfied since
\begin{align}
     2 c  := & \max_{k} \; \mathbb{E}\left[\left(\ln X_k \right)^2
\left(X_k +1\right)  \right] \notag \\
         \leq &  \left( s L_0   \max_{j} \; \left(\mathbb{E}\left[
\left(\ln \left| \Gamma_{j,L} \right| \right)^4 \right] \right)^{1/4}
             + \left(\mathbb{E}\left[\left(\ln Y_k
\right)^4 \right]\right)^{1/4} \right)^2     \notag \\
         & \qquad \times\left( \mathbb{E}\left[ X_k^2 + 2 X_k
+ 1 \right] \right)^{1/2}
          \leq s^2 L_0^2 \,  C \, .
\end{align}
The above result is based on the Cauchy-Schwarz inequality
and norm subadditivity.  Moreover, the last inequality
uses \eqref{eq:smombound} which also proves that
     the expectations of powers of logarithms of diagonal Green
functions are uniformely bounded by Lemma~\ref{lemma:momlog}.
In applying  Lemma~\ref{lemma:triv} it is also useful to note that
\begin{align}
     & \mathbb{E}\left[\ln \prod_{k= 0}^{N_x-1} X_k \right] - s \,
      \mathbb{E}\left( \ln \left| \big\langle \delta_0 , \big(
H_{\mathcal{T}_L} - E \big)^{-1} \delta_x \big\rangle \right|
          \right) \notag \\
     & = - \sum_{k=0}^{N_x-1} \mathbb{E}\left[\ln Y_k \right] -
\mathbb{E}\left[\ln R \right] \notag \\
     & \leq  N_x \max_{k} \left| \mathbb{E}\left[\ln Y_k \right]
\right| + s L_0 \max_{j}  \left| \mathbb{E}\left[ \ln \left|
\Gamma_{j,L} \right|\right]\right|
     \leq s \,C (N_x + L_0)  \, ,
\end{align}
where we have again used the fact that expectations of logarithms of
diagonal Green functions are uniformely bounded.

Summarizing the above estimates we obtain the bound
\begin{align}
     & \ln \mathbb{E}\left[ \left| \big\langle \delta_0 , \big(
H_{\mathcal{T}_L} - E \big)^{-1} \delta_x \big\rangle\right|^s\right]
- s \,
      \mathbb{E}\left( \ln \left| \big\langle \delta_0 , \big(
H_{\mathcal{T}_L} - E \big)^{-1} \delta_x \big\rangle \right|
          \right) \notag \\
     & \leq (N_x + 1) \ln C_s +  N_x s^2 L_0^2 \, C  + s \, C (N_x
+ L_0)  \notag \\
     & \leq \dist(0,x) \left(2 L_0^{-1} \ln C_s + s^2 L_0 \, C + 2
s \, C \right) \, , \label{eq:schluss}
\end{align}
where the last inequality holds provided $ \dist(0,x) \geq L_0 $.
Consequently, for a given $ \varepsilon > 0 $
we may then pick
$ L_0 = L_\varepsilon $ large enough and $ s_\varepsilon $ small
enough such that
the right side in \eqref{eq:schluss}
is smaller that $ \varepsilon  \, \dist(0,x) $ for every $ s \in
(0,s_\varepsilon) $.
\end{proof}

We close this subsection by compiling two elementary estimates on
expectations of functions of the diagonal of the Green function. The
first bounds concern fractional moments
of the Green function going back to \cite{AM}.
\begin{lemma}\label{lemma:ubound}
Assume {\bf A1} and let $ s \in (0, 1) $, $ z \in \mathbb{C} $ and $ 
L \in \mathbb{N} $. Then
\be
  \mathbb{E}\left[\left| \big\langle \delta_0 , \big(
H_{\mathcal{T}_{L}} - z \big)^{-1} \delta_0 \big\rangle \right|^{s}
\right]
     \leq C_s
\ee
and
\begin{equation}
     \mathbb{E}\left[\left| \big\langle \delta_0 , \big(
H_{\mathcal{T}_{L}} - z \big)^{-1} \delta_0 \big\rangle \right|^{-s}
\right]
     \leq \int_\mathbb{R} |\xi|^s \varrho(\xi) d\xi + |z|^s + K \, C_s \, ,
\end{equation}
where $ C_s $ is the constant appearing in \eqref{eq:smombound}..
\end{lemma}
\begin{proof}
The first inequality is an immediate consequence of the fractional
moment bound \eqref{eq:smombound}. The second one is a consequence of
the first and the
recursion relation which the
diagonal of the resolvent is well-known satisfy, cf.\ \cite{Kle98},
\begin{equation}
     \big\langle \delta_0 , \big( H_{\mathcal{T}_{L}} - z
\big)^{-1} \delta_0 \big\rangle
     = \Big( \omega_0 - z - \sum_{\dist(x,0) =1 } \big\langle
\delta_x , \big( H_{\mathcal{T}_{L}(x)} - z \big)^{-1} \delta_x
\big\rangle \Big)^{-1}
\end{equation}
where we recall that $\mathcal{T}_{L}(x) $ is that subtree of $
\mathcal{T}_L $ which is forward to $ x $.
\end{proof}
Lemma~\ref{lemma:ubound} in particular implies that any moment of the
logarithm of the Green function is uniformly bounded.
\begin{lemma}\label{lemma:momlog}
Assume {\bf A1} and let $ I \subset \mathbb{R} $ be a bounded Borel 
set and $ n \in
\mathbb{Z} $. Then
\begin{equation}
     \sup_{E \in I} \, \sup_{L\in \mathbb{N}} \;
\mathbb{E}\left[\, \left| \, \ln \left|
     \big\langle \delta_0 , \big( H_{\mathcal{T}_{L}} - E
\big)^{-1} \delta_0 \big\rangle \right| \right|^n\right]
     < \infty \, .
\end{equation}
\end{lemma}
\begin{proof}
This estimate immediately follows from Lemma~\ref{lemma:ubound}
and the fact that $ \left| \ln \xi \right| \leq \xi^\tau +
\xi^{-\tau} $ for any $ \xi > 0 $ and $ \tau \neq 0 $.
\end{proof}

\subsection{Lower bound on the Lyapunov exponent}\label{subsec:Lya}
In \cite{ASW05} we considered a Lyapunov exponent for the operator
$H_\mathcal{T} $ on the infinite regular rooted tree
with branching number $ K \geq 2 $, defining it as
\begin{equation}
     \gamma(z) := - \mathbb{E}\left[ \ln \left( \sqrt{K} \, \left|
\big\langle \delta_0 , \big( H_\mathcal{T} - z \big)^{-1} \delta_0
\big\rangle \right) \right| \right] \, .
\end{equation}
It was shown there (\cite[Thm.~3.1 \& Thm.~4.1]{ASW05}) that
\begin{enumerate}
     \item $ \gamma(z) $ is a positive harmonic function of $ z
\in \mathbb{C}^+ $
     and hence its boundary values $ \gamma(E+i0) $ with $ E \in
\mathbb{R} $ define a locally integrable function.
     \item For
all $ z \in \mathbb{C}^+ $ and all $ \alpha \in (0,1/2) $
\begin{equation}\label{eq:ASW}
     \gamma(z) \geq \frac{\alpha^2}{32 (K+1)^2} \left(
\delta\left( | \Gamma_0(z) |^{-2} , \alpha \right) \right)2
     \, ,
\end{equation}
where $ \Gamma_0(z) := \big\langle \delta_0 , \big( H_\mathcal{T} - z
\big)^{-1} \delta_0 \big\rangle $ and $\delta(\cdot,\cdot)$
is defined as follows.
\end{enumerate}

    \begin{definition}[Relative width]\label{def:relwitdth}
              For $ \alpha \in (0,1/2] $  the
     \emph{relative $ \alpha $-width} of a positive random variable $ X
$  is given by
              \begin{equation}\label{def:relwidth}
                      \delta(X,\alpha) := 1 - \frac{\xi_-(X,
\alpha)}{\xi_+(X, \alpha)}.
              \end{equation}
where $
\xi_-(X, \alpha) := \sup \{\, \xi \, , \mathbb{P}\left( X < \xi
\right) \leq \alpha \} $ and $ \xi_+ (X, \alpha) :=  \inf \{\,  \xi
\, ,   \mathbb{P}\left( X > \xi \right) \leq \alpha\} $.
\end{definition}

Our next task is to further estimate the right side of \eqref{eq:ASW}
from below. This will be done with the help of the
    following lemma.
\begin{lemma}\label{lemma:boundonrelwidth}
Let $ X $ be a positive random variable with probability measure $
\mathbb{P} $. Suppose
\begin{enumerate}
\item there exists $ \sigma \in (0,1] $ and $ C_\sigma < \infty $ such that
     $ \mathbb{P}\left( X \in I \right) \leq C_\sigma |I|^\sigma $
for all Borel sets $ I \subset [0, \infty) $ with $ | I | \leq 1 $.
\item there exists $ \tau > 0 $ such that  $ \mathbb{E}\left[ X^\tau
\right] < \infty $.
\end{enumerate}
Then for all $ \alpha \in (0,1/2) $
\begin{equation}
     \delta(X,\alpha) \geq \min\left\{1,
\left(\frac{1-2\alpha}{C_\sigma}\right)^{1/\sigma} \right\}
     \left(\frac{\alpha}{ \mathbb{E}\left[ X^\tau
\right]}\right)^{1/\tau} \, .
\end{equation}
\end{lemma}
\begin{proof}
The first assumption implies that $ 1 - 2 \alpha = \mathbb{P}\left\{
X \in \big(\xi_-(X,\alpha),\xi_+(X,\alpha)\big) \right\}
     \leq C_\sigma \big(\xi_+(X,\alpha)-\xi_-(X,\alpha)
\big)^\sigma $ provided $ \xi_+(X,\alpha)-\xi_-(X,\alpha) \leq 1 $.
   From the second assumption we conclude that
$ \alpha \leq  \mathbb{P}\left\{ X \in
\big(\xi_+(X,\alpha),\infty\big) \right\} \leq
\mathbb{E}\left[ X^\tau \right] /  \xi_+(X,\alpha)^{\tau} $ by a
Chebychev inequality.
Inserting these two estimates into \eqref{def:relwidth} completes the proof.
\end{proof}

\begin{lemma}\label{lemma:infpos}
Let  $ I \subset \mathbb{R} $ be a bounded Borel set and
\begin{equation}
    l(I)
     := \sup_{\alpha\in (0,1/2) } \frac{\alpha^2}{32 (K+1)^2}
\min\left\{1,\frac{1-2\alpha}{2 \, \|\varrho\|_\infty }\right\}
     \left(\frac{\alpha}{ \mathbb{E}\left[ \sup_{E \in I}
|\Gamma_0(E+i0) |^{-\tau}\right]}\right)^{2/\tau}
\end{equation}
where $ \tau $ is the constant appearing in Assumption~{\bf A1}. Then
$ \gamma(E+i0) \geq l(I) > 0 $ for any $ E \in I $.
\end{lemma}
\begin{proof}
In order to apply Lemma~\ref{lemma:boundonrelwidth} to the right side
in \eqref{eq:ASW} we need to check its assumptions. We first note that
by the  Krein formula $| \Gamma_0(z) |^{-2} = (\omega_0 - a)^2 + b^2
$ with suitable $ a $, $ b \in \mathbb{R} $. An elementary
computation shows that
for every Borel set $ I \subset [0, \infty) $ with $ | I | \leq 1 $
\begin{equation}
       \int_\mathbb{R} \varrho(\xi)
     \indfct_{\{ (\xi - a)^2 + b^2 \in I \} } d\xi
     \leq \int_I \frac{\|\varrho\|_\infty}{\sqrt{\xi - b2 }}
\indfct_{\{\xi \geq b^2\}} d \xi
      \leq 2 \|\varrho\|_\infty \sqrt{|I|} \, .
\end{equation}
Moreover, Lemma~\ref{lemma:ubound} guarantees that $
\sup_{E \in I} \mathbb{E}\left[ |\Gamma_0(E+i0) |^{-\tau}\right] <
\infty $.
\end{proof}
Associated with $ \gamma(z) $ is the following finite-volume approximation
\begin{equation}\label{def:gl}
     \gamma_L(z) := -  \mathbb{E}\left[\ln \left( \sqrt{K} \,
     \left| \big\langle \delta_0 , \big( H_{\mathcal{T}_L} - z
\big)^{-1} \delta_0 \right| \right) \right] \, .
\end{equation}
It is easy to see that $ \gamma_L(z) $ also defines a harmonic
function of $ z \in \mathbb{C}^+ $. Moreover, its boundary values $
\gamma_L(E) $ are defined
everywhere by setting $ z = E \in \mathbb{R} $ in \eqref{def:gl}.
Strong resolvent convergence implies that $ \lim_{L \to \infty}
\gamma_L(z) = \gamma(z) $ for every $ z \in \mathbb{C}^+ $.
Assumption~{\bf A2} guarantees that
this convergence
holds and is locally uniform also for real arguments.
\begin{lemma}\label{lemma:convgamma}
Suppose {\bf A2} holds for a bounded Borel set $ I \subset \mathbb{R} $. Then
\begin{equation}\label{eq:unconv} \lim_{L \to \infty}  \, \sup_{E \in
I } \, \left| \gamma_L(E) - \gamma(E+i0) \right| = 0\, .
\end{equation}
\end{lemma}
\begin{proof}
     Since $ \gamma_L(E)  $ are uniformly bounded for $ E \in I $,
cf.\ Lemma~\ref{lemma:momlog}.
     By the Arzela-Ascoli theorem Assumption~{\bf A2} thus implies that
     every subsequence of $ \gamma_L $ has a uniformly convergent
subsequence.
     The claim \eqref{eq:unconv}  then follows by showing that any
pointwise limit of $ \gamma_L(E) $ coincides with $ \gamma(E+i0) $.
     This is derived from the above mentioned strong resolvent
convergence and the dominated convergence theorem, which imply that
     for any bounded and compactly supported function
     $ \phi \in L^\infty_c(\mathbb{R}) $
     \begin{equation}
        \int_\mathbb{R}\gamma(E+i0) \phi(E) dE = \lim_{L \to
\infty} \int_\mathbb{R} \gamma_L(E) \phi(E)  dE  = \int_\mathbb{R}
\lim_{L\to \infty} \gamma_L(E) \phi(E) dE \, .
     \end{equation}
     provided $ \lim_{L\to \infty} \gamma_L(E) $ exists for
Lebesgue-almost all $ E \in \mathbb{R} $.
\end{proof}

\subsection{Proof of Theorem~\ref{lemma:decGreen}}

\begin{proof}[Proof of Theorem~\ref{lemma:decGreen}]
If $ x \neq y $ is in the future of $ y $, the Green function
factorizes according to
\begin{equation}
     \big\langle \delta_y , \big( H_{\mathcal{T}_L} - E \big)^{-1}
\delta_x \big\rangle =
     \big\langle \delta_y , \big( H_{\mathcal{T}_L} - E \big)^{-1}
\delta_y \big\rangle
     \big\langle \delta_v , \big( H_{\mathcal{T}_{L}(v)} - E
\big)^{-1} \delta_x \big\rangle
\end{equation}
where $ v $ is that forward neighbor of $ y $ which lies on the
unique path connecting $ x $ and $ y $.
We  may therefore suppose without loss of generality that $ y $
coincides with the root in $\mathcal{T}_L $.

In this case, Lemma~\ref{lemma:decGreen1} bounds the fractional
moment of the Green function by an exponential involving
\begin{equation}
      \mathbb{E}\left[ \ln \big\langle \delta_y , \big(
H_{\mathcal{T}_L} - E \big)^{-1} \delta_x \big\rangle \right]
     = - \sum_{j=0}^{\dist(y,x)} \left( \gamma_{L-j}(E) + \ln
\sqrt{K} \right) \, ,
\end{equation}
where we the last equality results from \eqref{eq:Gfactor},
stationarity and the definition of the finite-volume
Lyapunov exponent in~\eqref{def:gl}.
According to Lemma~\ref{lemma:convgamma}, for a given $
\varepsilon > 0 $ there exists $ L_\varepsilon \in \mathbb{N} $
such that $ \gamma_{L}(E) \geq \gamma(E+i0) - \varepsilon \geq
l(I) - \varepsilon $ for all $ E \in I $ and $ L \geq
L_\varepsilon $, where
$l(I) > 0$ was defined in Lemma~\ref{lemma:infpos}.
The proof is completed by choosing $ \varepsilon $ small enough in
the last estimate and in Lemma~\ref{lemma:decGreen1}.
\end{proof}

\section{Proof of Poisson statistics for tree operators}\label{sec:Proofmain}

We will follow the general strategy outlined in 
Subsection~\ref{sec:sufficient}: Poisson statistics are established through the proof of the negligibility in probability  condition~\eqref{eq:condition}, which holds due to the divergence of fluctuations of the value at a fixed site of the normalized eigenfunctions.    

\subsection{Divergent fluctuations of the spectral measure in the 
bulk}\label{Subsec:deloc}

The following theorem allows to conclude that for any $ x \in 
\mathcal{T} $ the sequence of
spectral measures $ \{ \sigma_{x,L} \} $ has divergent fluctuations
in the sense of Definition~\ref{def:unif},
at any $ E \in \mathbb{R} $ for which the average over the disorder of theses
measures is non zero.

\begin{theorem}[Negligibility in probability]\label{thm:deloc}
Assume {\bf A1} and {\bf A2 } holds for a bounded Borel set $ I 
\subset \mathbb{R} $.
If $ I_L \subset I $ are bounded Borel sets such that $
\limsup_{L\to \infty} | I_L| | \mathcal{T}_L | < \infty $, then for
any $ x \in \mathcal{T} $
\begin{equation}\label{eq:deloc}
     \Plim_{L\to \infty} | \mathcal{T}_L | \;\sigma_{x,L}(I_L) = 0 \, .
\end{equation}
\end{theorem}
For the proof we now fix $ x \in \mathcal{T} $, and for $ L \in \mathbb{N} $
large enough so that $ x \in\mathcal{T}_L $ and
every $ y \in \mathcal{T}_L $ we define the ratio
\begin{equation}
     g_{y,L}(E) := \frac{\left| \big\langle \delta_x , \big(
H_{\mathcal{T}_L} -E \big)^{-1} \delta_y \big\rangle \right|^2}{
     \big\langle \delta_y , \big( H_{\mathcal{T}_L} -E \big)^{-2}
\delta_y \big\rangle} \, .
\end{equation}
It is well-defined for Lebesgue-almost all $ E \in \mathbb{R} $.
Moreover, by the rank-one pertubation formula and the spectral
theorem it is seen to
enjoys the following properties:
\begin{enumerate}
     \item  $ g_{y,L}(E) $ is independent of the value of the
potential at $ y \in \mathcal{T}_L $.
     \item The function $   E \mapsto g_{y,L}(E) $ has a
continuous extension on $ \mathbb{R} $. Moreover,
     if the eigenvalue $ E_n(\mathcal{T}_L) $ of $
H_{\mathcal{T}_L} $ is non-degenerate, then the corresponding
eigenfunction satisfies
     \begin{equation}\label{eq:psig}
         \left|  \big\langle \delta_x, \psi_{n}(\mathcal{T}_L)
\big\rangle \right|^2 = \lim_{E \to E_n(\mathcal{T}_L)} g_{y,L}(E)
     \end{equation}
\end{enumerate}

Theorem~\ref{thm:deloc} will now be a consequence of the
following result.
\begin{lemma} Under the assumptions of Theorem~\ref{thm:deloc}
for any $ \varepsilon > 0 $
\begin{equation}\label{eq:eigesumnnulla}
\lim_{L\to \infty}
     \sum_{y \in \mathcal{T}_L} \mathbb{E}\left[
\sigma_{y,L}\big(E \in I_L \, : \, g_{y,L}(E) \geq \varepsilon \,
|\mathcal{T}_L|^{-1} \big) \right] = 0 \, .
\end{equation}
\end{lemma}
\begin{proof}
The proof is based on the spectral averaging principle (cf.\
\cite{SiWo86,CaLa90}),
\begin{equation}\label{eq:specav}
     \int_\mathbb{R} \varrho(\omega_y)\,  \sigma_{y,L}(I)
d\omega_y \leq \| \varrho \|_\infty \int_\mathbb{R} \sigma_{y,L}(I)
\, d\omega_y
     \leq \| \varrho \|_\infty | I | \,
\end{equation}
for all bounded Borel sets $ I \subset \mathbb{R} $.
Using this inequality and the fact that $ g_{y,L}(E) $ does not
depend on $ \omega_y $,
the prelimit in \eqref{eq:eigesumnnulla} can be bounded
from above by
\begin{equation}\label{eq:sumspecav}
     \| \varrho \|_\infty \, \sum_{y \in \mathcal{T}_L}
\int_{I_L}  \mathbb{P}\left(  g_{y,L}(E) \geq \varepsilon \,
|\mathcal{T}_L|^{-1}  \right) dE \, .
\end{equation}
We now pick $ N \in \mathbb{N} $ and split the summation in
\eqref{eq:sumspecav} into two terms. The first term
collects all contributions corresponding to $ \mathcal{T}_{L-N}
\subset \mathcal{T}_L $,
\begin{equation}
         \sum_{y\in \mathcal{T}_{L-N}} \int_{I_L}
\mathbb{P}\left(      g_{y,L}(E) \geq \varepsilon \,
|\mathcal{T}_L|^{-1}  \right) dE
     \leq \big| \mathcal{T}_{L-N} \big| \, \big| I_L \big| \, .
\end{equation}
In the limit $ L \to \infty $ this term is arbitrarily small for $ N
$ large enough.
To estimate the remaining second term, we abbreviate
$ \alpha_{y,L}(E) := \big\langle \delta_y , \big( H_{\mathcal{T}_L}
-E \big)^{-2} \delta_y \big\rangle  $ and write
\begin{multline}\label{eq:sectermeigen}
      \mathbb{P}\left(  g_{y,L}(E) \geq \varepsilon \,
|\mathcal{T}_L|^{-1}  \right)  =
     \mathbb{P}\left( |\mathcal{T}_L | \left| \big\langle \delta_x
, \big( H_{\mathcal{T}_L} -E \big)^{-1}
     \delta_y \big\rangle \right|^2 \geq \varepsilon  \,
\alpha_{y,L}(E) \right) \\
     \leq \mathbb{P}\left( |\mathcal{T}_L | \left| \big\langle
\delta_x , \big( H_{\mathcal{T}_L} -E \big)^{-1}
     \delta_y \big\rangle \right|^2 \geq \varepsilon \alpha
\right) + \mathbb{P}\left(\alpha_{y,L}(E) < \alpha \right)
\end{multline}
where the last inequality holds for any $ \alpha \in (0, \infty) $.
The first term on the right side of \eqref{eq:sectermeigen} gives
rise to the following contribution to the sum in \eqref{eq:sumspecav},
\begin{multline}
\sum_{y\in \mathcal{T}_L \backslash \mathcal{T}_{L-N}} \int_{I_L}
\mathbb{P}\left( |\mathcal{T}_L | \left| \big\langle \delta_x , \big(
H_{\mathcal{T}_L} -E \big)^{-1}
     \delta_y \big\rangle \right|^2 \geq \varepsilon \alpha \right) dE
     \leq \frac{|\mathcal{T}_L |^s}{\varepsilon^s \alpha^s K^{s(L-N)}} \\
     \times \sup_{E \in I } \; | I_L | \mkern-5mu
     \sum_{y\in \mathcal{T}_L \backslash \mathcal{T}_{L-N}}
\mkern-8mu K^{s\dist(x,y)} \,
     \mathbb{E}\left[ \left| \big\langle \delta_x , \big(
H_{\mathcal{T}_L} -E \big)^{-1}
     \delta_y \big\rangle \right|^{2s} \right] \, , \label{eq:letzes}
\end{multline}
where $ I \subset \mathbb{R} $ is some bounded Borel set which
contains eventually all $ I_L $.
While the prefactors on the right side of \eqref{eq:letzes} remains
finite in the limit $ L \to \infty $,
the supremum
converges to zero in this
limit, since it is bounded by
$ | I_L | | \mathcal{T}_L |\, C \exp\left(- 2s \delta\,  (L-N)
\right) $ for sufficently
small $ s $ by Theorem~\ref{lemma:decGreen}.
To complete we note that the second term in \eqref{eq:sectermeigen},
converges to zero as
$ \alpha \downarrow 0 $, uniformly in $ E \in I $, $ L \in \mathbb{N}
$ and $ y \in \mathcal{T}_L $.
This follows from the bound
\begin{align}
\mathbb{P}\left(\alpha_{y,L}(E) < \alpha \right) & \leq
\mathbb{P}\left(\left\| \big( H_{\mathcal{T}_L} - E \big) \delta_y
\right\|^{-2} < \alpha \right) \notag \\
     & \leq \alpha^s \left(\big\| \left( A +
B - E \right) \delta_y \big\|^{2s}
     + \mathbb{E}\left[|\omega_y|^{2s}\right] \right)
     \, ,
\end{align}
where the last step requires $ 2s < \min(1,\tau) $.
\end{proof}

\begin{proof}[Proof of Theorem~\ref{thm:deloc}]
Wegner's bound \eqref{eq:Wegner} implies
that $ I_L $ carries only a finite number of eigenvalues
\begin{equation}\label{eq:Wegcons}
     \lim_{N \to \infty} \sup_{L \in \mathbb{N}} \;
\mathbb{P}\left( \Tr P_{I_L}(H_{\mathcal{T}_L} ) \geq N \right) = 0
\, .
\end{equation}
It therefore remains to prove that for any $ \varepsilon >0 $
\begin{equation}\label{eq:eigesumnnull}
\lim_{L \to 0 } \; \mathbb{E}\left[ \sum_{E_n(\mathcal{T}_L) \in I_L}
\indfct\left\{\left|  \big\langle \delta_x, \psi_{n}(\mathcal{T}_L)
\big\rangle \right|^2
     \geq \varepsilon \, |\mathcal{T}_L|^{-1} \right\} \right] = 0 \, ,
\end{equation}
where $ \indfct\{\cdots\} $ stands for the indicator function.
Using the fact that $ H_{\mathcal{T}_L} $ has almost surely no
degenerate eigenfunctions (cf.\ Proposition~\ref{prop:W+M})
and \eqref{eq:psig},
the left side in \eqref{eq:eigesumnnull} is seen to be equal
to the left side in \eqref{eq:eigesumnnulla}.
\end{proof}

Theorem~\ref{prop:infdiv} now guarantees that any accumulation point 
of $\big\{\mu_{L}^E
\big\} $ is a random Poisson measure. The uniqueness of the accumulation point
will be proven by establishing uniqueness of the intensity measure, which
is defined next.

\subsection{The intensity measure}\label{subsec:intensity}

For a random point measure $ \mu $  the intensity measure is
defined as the average
\begin{equation}
     \overline{\mu} := \mathbb{E}\left[ \mu  \right] \, .
\end{equation}
Thus $ \overline{\mu}_{L}^E$  are the intensity measures of
the random point measures $ \mu_{L}^E $.  We shall also use the symbol
$ \overline{\mu}^E $  for the intensity measure of a given 
accumulation point of the sequence  $\big\{\mu_{L}^E \big\} $.  

Let us proceed with a more explicit representation for $
\overline{\mu}_{L}^E $.
For any Borel set $I \subset \mathbb{R}  $ we have
\begin{align}
         \overline{\mu}_{L}^E(I ) & =  \mathbb{E}\left[  \Tr
P_{E+ I/ |\mathcal{T}_L| }(H_{\mathcal{T}_L})\right] \notag \\
         & =   \sum_{x \in \mathcal{T}_L}
              \mathbb{E}\left[ \big\langle \delta_x ,
P_{E+ I/ |\mathcal{T}_L| }(H_{\mathcal{T}_L}) \, \delta_x \big\rangle
\right] \notag \\
         & =  \sum_{n=0}^{L} \, K^{L-n} \;
\mathbb{E}\left[ \big\langle \delta_{x_n}
,  P_{E+I \, |\mathcal{T}_L|^{-1}}(H_{\mathcal{T}_L}) \, \delta_{x_n}
\big\rangle \right]
\, , \label{eq:calc}
\end{align}
where we used the fact that the expectation in the second line does
not depend on $ x $ as long as $ \dist(0,x) $ is constant. Moreover,
$ x_n $ denotes any vertex with
$\dist(x_n, \partial \mathcal{T}_L ) = n $.
In view of Lemma~\ref{lemma:dos} in the appendix, the above
calculation~\eqref{eq:calc} suggests that the intensity measure $
\overline{\mu}_{L}^E $
converges for Lebesgue almost
all $ E \in \mathbb{R} $ to Lebesgue measure times the canopy density
of states given by
\begin{equation}\label{eq:rep}
      d_\mathcal{C}(E) := \frac{K-1}{K} \sum_{n=0}^\infty K^{-n}
\, \pi^{-1} \, \mathbb{E}\left[ \Im \,
\big\langle \delta_{x_n} , (H_{\mathcal{C}} - E -i0)^{-1} \delta_{x_n}
\big\rangle \right] \, .
\end{equation}

\begin{theorem}[Limiting intensity measure]\label{thm:intensity}
Under  assumption~{\bf A1} and {\bf A2} (or alternatively {\bf A1'} below) 
for Lebesgue-almost all $ E \in \mathbb{R} 
$ the intensity measure $
\overline{\mu}^E  $ of
any weak accumulation point $ \mu^E$ of the sequence $ \mu^E_L  $ is given by
\begin{equation}\label{accpoint}
     \overline{\mu}^E(I) = \lim_{L\to \infty}
\overline{\mu}^E_L(I) = d_\mathcal{C}(E) \, |I|
\end{equation}
for all bounded Borel sets $ I \subset \mathbb{R} $.
\end{theorem}
\begin{proof}
As an immediate consequence of Wegner's
estimate~\eqref{eq:Wegner} and the first line in \eqref{eq:calc} we
have that for Lebesgue-almost all $ E \in \mathbb{R} $ and all $ L \in
\mathbb{N} $ the measures $  \overline{\mu}^E_L $ are absolutely
continuous with bounded density,
\begin{equation}\label{eq:boundden}
      \frac{\overline{\mu}^E_L(d\xi)}{d\xi} \leq \| \varrho \|_\infty \, .
\end{equation}
The same applies to any accumulation point $  \overline{\mu}^E $.
As a consequence, the linear functional given by
$ \overline{\mu}_{L}^E(\psi) := \int_\mathbb{R} \psi(\xi)\,
\overline{\mu}^E_L(d\xi) $
is uniformly equicontinuous on the space of non-negative integrable
functions on the real line,
$ \psi \in L^1_+(\mathbb{R})  $. More precisely, \eqref{eq:boundden} yields
\begin{equation}\label{eq:equicont}
\left| \overline{\mu}_{L}^E(\phi)- \overline{\mu}_{L}^E(\psi) \right|
     \leq \|\varrho\|_\infty \, \left\| \phi - \psi \right\|_1
\end{equation}
for all $ \phi $, $\psi \in L^1_+(\mathbb{R})  $. Using this and the 
fact that the functions
$ \varphi_z := \pi^{-1} \, \Im (\cdot - z )^{-1} $ with $ z \in
\mathbb{C}^+ $ are dense in $ L^1_+(\mathbb{R}) $ implies
that it suffices to check \eqref{accpoint} if the indicator function of $ I $
is replaced by $ \varphi_z $.

Moreover, elementary considerations show that it suffices to verify
\begin{equation}
     \lim_{L\to \infty} \int_{\mathbb{R}} \left|
\overline{\mu}^E_L\big(\varphi_z\big) - d(E)  \, \left\| \varphi_z \right\|_1
  \right| d E = 0 \,
\end{equation}
with $ z \in \mathbb{C}^+ $ fixed but arbitrary.
A computation similiar to  \eqref{eq:calc} and the fact that $ 
\left\| \varphi_z \right\|_1 = 1 $ then proves that this derives from
\begin{multline}\label{eq:intcl}
     \lim_{L \to \infty} \int_{\mathbb{R}} \mathbb{E}\left[
     \left|
     \Im  \big\langle  \delta_x , \big(H_{\mathcal{T}_L} - E - z
\, |\mathcal{T}_L|^{-1} \big)^{-1}
     \delta_x \big\rangle \right.\right. \\
     \left.\left. -
     \Im \, \big\langle \delta_x , (H_{\mathcal{C}} - E -i0)^{-1}
\delta_x \big\rangle \right| \right] d E = 0
\end{multline}
for $ x \in \mathcal{C} $ with $ \dist(x, \partial \mathcal{C}) \in
\mathbb{N}_0 $ fixed but arbitrary.

For a proof of \eqref{eq:intcl}, we appeal to Riesz's theorem which guarantees that the claimed $ L^1 $-convergence follows
from the almost
sure convergence of the integrand in \eqref{eq:intcl}  with respect to the product of the probability measure and
Lebesgue measure, and the equality of the integrals  
\begin{multline}
     \lim_{L\to \infty} \frac{1}{\pi} \, \int_{\mathbb{R}}
     \mathbb{E} \left[ \Im   \big\langle \delta_x , \big(H_{\mathcal{T}_L} - E
- z \, |\mathcal{T}_L|^{-1} \big)^{-1}
     \delta_x \big\rangle \right] \, dE  \\
     = \frac{1}{\pi} \, \int_{\mathbb{R}} \mathbb{E}\left[ \Im \,
\big\langle \delta_{x_n} , (H_{\mathcal{C}} - E -i0)^{-1} \delta_{x_n}
\big\rangle \right] \, dE = 1 \, . 
\end{multline}
In fact, we only need to show that the integrand in \eqref{eq:intcl}
converges in distribution with respect to the product measure.
To prove the latter we  
first note that one has the non-tangential limit
\begin{equation}
         \lim_{L \to \infty} \, \big\langle \delta_x ,
(H_{\mathcal{C}} - E - z \, |\mathcal{T}_L|^{-1})^{-1} \delta_x \big\rangle
     = \big\langle \delta_x, (H_{\mathcal{C}} - E -i0)^{-1}
\delta_x \big\rangle
\end{equation}
for Lebesgue-almost all $ E \in \mathbb{R} $.
Moreover, using the resolvent identity twice, we obtain the inequality
\begin{multline}
     \left| \big\langle  \delta_x , \big(H_{\mathcal{T}_L} - E - z
\, |\mathcal{T}_L|^{-1} \big)^{-1} \delta_x \big\rangle -
         \big\langle  \delta_x , \big(H_{\mathcal{C}} - E - z
\, |\mathcal{T}_L|^{-1} \big)^{-1} \delta_x \big\rangle \right| \\
     \leq   \left| \big\langle  \delta_x, \big(H_{\mathcal{T}_L} -
E - z |\mathcal{T}_L|^{-1} \big)^{-1}  \delta_{0_{L}} \big\rangle
      \big\langle  \delta_{0_{L}}, \big(H_{\mathcal{T}_L} - E - z
\, |\mathcal{T}_L|^{-1} \big)^{-1} \delta_x \big\rangle \right| \\
         \times
         \left| \big\langle \delta_{0^-_{L}},
\big(H_{\mathcal{C}} - E - z \, |\mathcal{T}_L|^{-1} \big)^{-1}
\delta_{0^-_{L}}  \big\rangle \right| \, ,
     \label{eq:resid}
\end{multline}
where $ 0_L $ is the root in $ \mathcal{T}_L $ and $0^-_{L} $ is its
backward neighbor.
The right side converges to zero in distribution with respect to the
product of the probability measure and
     Lebesgue measure  on any bounded interval. This follows from Theorem~\ref{lemma:decGreen} (or alternatively 
Proposition~\ref{prop:DKS} below)
and the fact that the factional-moment bound
\eqref{eq:smombound} implies that
     the probability that the last term in \eqref{eq:resid} is
large is bounded.
\end{proof}

\subsection{Proof of Theorem~\ref{thm:main}}\label{Subsec:Proof}
Theorem~\ref{thm:main} may be stated using the characterisation of
the Poisson process
in terms of  its characteristic functional. Namely, the random measure $
\mu^E $ is Poisson if
for any bounded Borel set $ I \subset \mathbb{R} $ and $ t \geq 0 $
\begin{equation}\label{eq:charfunc}
    \mathbb{E}\left[e^{- t \mu^E(I)}\right] = \exp\left( -
\mathbb{E}\left[\mu^E(I)\right] \left(1 - e^{-t}\right) \right) \, .
\end{equation}
Given Theorem~\ref{prop:infdiv} and Theorem~\ref{thm:deloc}, the proof of
\eqref{eq:charfunc} is
basically a repetition of well-known arguments how to conclude the
Poisson nature of accumulation
points from
infinite divisibility and the exclusion of double points \cite{Kal02}.
\begin{proof}[Proof of Theorem~\ref{thm:main}]
Let $ \mu^E $ be an accumulation point of $ \{ \mu_L^E \} $.
Theorem~\ref{prop:infdiv} implies that for any $ N \in \mathbb{N} $
and any bounded Borel set $ I \subset \mathbb{R} $
\begin{align}\label{eq:weakconv}
       \lim_{L \to \infty} \; \mathbb{E}\left[\,
\exp\left(-t \sum_{\dist(0,x)=N} \,\mu_{x,L}^{E}(I)\right)\right]  & \notag \\
     = \lim_{L \to \infty} \mathbb{E}\left[e^{-t \mu_L^E(I)}\right]
     & =  \mathbb{E}\left[e^{-t \mu^E(I)}\right] \, .
\end{align}
Since the measures in the left side of \eqref{eq:weakconv} are iid,
the expectation factorizes into a $ K^N $-fold product of
\begin{align}
     \mathbb{E}\left[ e^{- t \mu_{x,L}^N(I) } \right] & =
\sum_{m=0}^\infty e^{-t m} \; \mathbb{P}\left(\mu_{x,L}^{E}(I) = m
\right) \notag\\
         & = 1 - \mathbb{E}\left[\mu_{x,L}^{E}(I) \right]
\left(1 - e^{-t}\right) + R_{x,L}(I) \, , \label{eq:ppppp}
\end{align}
where
\begin{align}
      0 \leq  R_{x,L}(I) := &
\sum_{m=2}^\infty\mathbb{P}\left(\mu_{x,L}^{E}(I) = m \right)  \left[
m \left( 1- e^{-t}\right) + e^{-tm} -1 \right]  \notag \\
     \leq &\;   \sum_{m=2}^\infty (m-1) \;
\mathbb{P}\left(\mu_{x,L}^{E}(I) = m \right)
     =     \sum_{m = 2}^\infty \mathbb{P}\left(
\mu_{x,L}^{E}(I) \geq m  \right)
\end{align}
By \eqref{prop:minami} this term is arbitrarily small in the limit $ L
\to \infty $ provided $ N $ is large enough.
The second term in \eqref{eq:ppppp} converges,
\begin{equation}
     \lim_{L \to \infty}\,  K^N \,
\mathbb{E}\left[\mu_{x,L}^{E}(I) \right] =
\mathbb{E}\left[\mu^E(I)\right] \, .
\end{equation}
The claim now follows by taking the subsequent limit $ N \to \infty $
in \eqref{eq:weakconv} from the fact that
$ \lim_{n\to\infty} \big(1+ x_n/n)^n = e^x $ for any complex-valued
sequence with $ \lim_{n\to \infty} x_n = x $.
\end{proof}

\section{Complete localization for random operators on the canopy graph}
\label{subsec:Loccan}

We shall now prove Theorem~\ref{thm:loc}, which asserts that, under 
assumptions  {\bf A1} and {\bf A2},
on the canopy graph the random Schr\"odinger operator has only pure 
point spectrum, i.e., a complete set of square integrable 
eigenfunctions.
The argument is based on the Simon-Wolff criterion \cite{SiWo86}, for 
which a sufficient condition is that for every energy the Green 
function be almost surely square summable, when summed over one of 
its  arguments.   (Through spectral averaging a.s. properties of the 
Green function areshared by the  eigenfunctions.)

Applying the above criterion, an intuitive reason for localization on 
the canopy graph is that the number of points at distance $n$ from 
$x_0$ grows there as $K^{n/2}$, which is square root of the 
corresponding number for the regular tree.  By 
Theorem~\ref{lemma:decGreen}, the  Green function decays at a rate 
which - if one could ignore large deviations, would yield square 
summability.   That in itself is not enough since typically the sum 
of the Green function is much larger that the sum of the typical 
values -- otherwise, the result would be valid also for the full 
homogeneous tree.  While the corresponding statement is not valid in 
that case, square summability is missed there rather marginally. 
Thus it may be not that surprising that the significant reduction in 
the number of sites suffices to yield the required summability.   We 
establish that with the help of a fractional moment estimate, and 
making use of the two lemmas which follow.\\

We now regard $ \mathcal{T}_L $ as being embedded into $
\mathcal{C} $
in such a way that the outer boundary $ \partial \mathcal{T}_L $ is
embedded into $ \partial \mathcal{C} $
for every $ L \in \mathbb{N} $.
\begin{lemma}
Assume {\bf A1} and {\bf A2 } holds for a bounded Borel set $ I 
\subset \mathbb{R} $. Then there
exists $ s \in (0,1) $ such that for all $ x \in \mathcal{C} $ and
Lebesgue-almost all $ E \in I $
\begin{equation}\label{eq:finits}
       \sup_{ \eta \neq 0} \sup_{L \geq L_x} \;
      \mathbb{E}\left[ \big\langle \delta_x , \big|
H_{\mathcal{T}_L} - E -i \eta \big|^{-2} \delta_x \big\rangle^s
\right] < \infty \, ,
\end{equation}
where  $ L_x := \min \{ L \in \mathbb{N} \, : \,  x \in \mathcal{T}_L \} $.
\end{lemma}
\begin{proof}
We first note that the inequality
\begin{equation}
      \big\langle \delta_x , \big| H_{\mathcal{T}_L} - E - i \eta
\big|^{-2} \delta_x \big\rangle
\leq \big\langle \delta_x , \big| H_{\mathcal{T}_L} - E \big|^{-2}
\delta_x \big\rangle
\end{equation}
implies that we only need to bound
the $ \ell^2(\mathcal{T}_L) $-norm in \eqref{eq:finits} for $ \eta =
0 $. The expectation of the fractional-moment of this $
\ell^2(\mathcal{T}_L) $-norm is split
into two contributions. One involves all terms corresponding to the
finite subtree
\be\label{eq:subtreedef}
     \mathcal{C}(x) := \left\{ y \in \mathcal{C} \, : \,  y  \;
\mbox{is forward (in the direction of $ \partial \mathcal{C} $) or
equal to} \; x \right\} \, ,
\ee
which has $ x $ as its root, and the other collects all remaining
terms.
Employing the elementary inequality $ ( \sum_j \alpha_j )^s \leq
\sum_j \alpha_j^s $, which is valid for any $ s \in (0,1) $ and any
collection
of non-negative numbers $ \alpha_j $, we thus obtain
\begin{align}
     & \mkern-150mu \mathbb{E}\left[ \left(\sum_{y \in
\mathcal{T}_L} \left|\big\langle \delta_x , \big( H_{\mathcal{T}_L} -
E \big)^{-1} \delta_y \big\rangle\right|^2\right)^s \right] \leq S_1
+ S_2 \, ,
     \label{eq:smomauft}\\
     \mbox{where} \qquad &  S_1 := \sum_{y \in\mathcal{C}(x)}
\mathbb{E}\left[ \left|\big\langle \delta_x , \big( H_{\mathcal{T}_L}
- E \big)^{-1} \delta_y \big\rangle\right|^{2s} \right] \notag \\
     &  S_2:= \mathbb{E}\left[ \left(\sum_{y \in
\mathcal{T}_L\setminus\mathcal{C}(x)} \left|\big\langle \delta_x ,
\big( H_{\mathcal{T}_L} - E \big)^{-1} \delta_y
\big\rangle\right|^2\right)^s \right] \notag \, .
\end{align}
By the fractional-moment bound \eqref{eq:smombound} the first terms,
$S_1$, is bounded for any $ s \in (0,1/2) $ by a constant, $
|\mathcal{C}(x)| C $, which is independent of $ L \geq L_x $ and $ z
\in \mathbb{C}^+ $.

To bound the second term, $ S_2 $, we use the fact that
the Green's function factorizes,
\begin{equation}
     \big\langle \delta_x , \big( H_{\mathcal{T}_L} - E \big)^{-1}
\delta_y \big\rangle
     = \big\langle  \delta_x , \big( H_{\mathcal{T}_L} - E
\big)^{-1} \delta_{v} \big\rangle \,
     \big\langle  \delta_w ,  \big( H_{\mathcal{C}(w)} - E
\big)^{-1} \delta_y \big\rangle \, ,
\end{equation}
where $ v $ is the first joint ancestor of $ x $ and $ y $, and $ w $
is that neighbor of $ v $ which has the least distance from $ y $.
We may therefore organize the summation in $ S_2 $ as follows.
We sum over the vertices on the unique path in
$ \mathcal{P}(x) \subset \mathcal{C} $ which
connects $ x $ and ``infinity'', cf.\ Figure~\ref{Fig:Canopy}. For
each vertex along this path we
then collect terms of the form
\begin{equation}
     S(w)
     := \sum_{y \in \mathcal{C}(w)}
\left| \big\langle  \delta_w ,  \big( H_{\mathcal{C}(w)} - E
\big)^{-1} \delta_y \big\rangle \right|^2 \, ,
\end{equation}
which stem from the $ K-1 $ neighbors $ w $ of $ v $, which are not
in  $ \mathcal{P}(x) $.
Consequently, the second term in \eqref{eq:smomauft} is bounded according to
\begin{align}\label{eq:smomauft1}
     S_2 & \leq \sum_{v \in  \mathcal{P}(x) \cap \mathcal{T}_L}
     \mathbb{E}\Big[  \left| \big\langle  \delta_x , \big(
		H_{\mathcal{T}_L} - E \big)^{-1} \delta_{v} 
\big\rangle \right|^{2s} \Big[ 1 +
	\sum_{\substack{\dist(w,v) = 1 \\ w \not\in \mathcal{P}(x)}} 
S(w)^s \Big] \Big] \notag \\
     &\leq \sum_{v \in  \mathcal{P}(x) \cap \mathcal{T}_L} 
	\sum_{\substack{\dist(w,v) = 1 \\ w \not\in \mathcal{P}(x)}}
	\mathbb{E}\left[  \left| \big\langle  \delta_x , \big(
		H_{\mathcal{T}_L} - E \big)^{-1} \delta_{v} 
\big\rangle \right|^{4s} \right]^{1/2}
	\left[ 1 + \left(\mathbb{E}\left[S(w)^{2s}\right]^{1/2} 
\right) \right]\notag \\
	&\leq C(s,K)  \!\!\!\sum_{v \in  \mathcal{P}(x) \cap
\mathcal{T}_L} \!\!\left(\mathbb{E}\left[ K^{2s \dist(x,v)} \,
         \left| \big\langle  \delta_x , \big(
H_{\mathcal{T}_L} - E \big)^{-1} \delta_{v} \big\rangle \right|^{4s}
\right]\right)^{1/2} \notag \\
     & \mkern250mu \times      	\left[ 1 +  \left(\mathbb{E}\left[ 
\frac{1}{|C(w)|^{2s}} \,
S(w)^{2s}  \right]\right)^{1/2} \right]\, ,
\end{align}
where $ C(s,K) < \infty $ is
independent of $ w $ and $ v $, and $ w  $
is any of the $ (K-1) $ neighbors of $ v $ with $ w \notin \mathcal{P}(x) $.
According to Lemma~\ref{lemma:maxdel} below, the last term in the
right side of \eqref{eq:smomauft1} is bounded from above
by a constant which is independent of $ w $.
Lemma~\ref{lemma:decGreen} then proves that the remaining sum over $
v \in   \mathcal{P}(x) \cap \mathcal{T}_L $
in \eqref{eq:smomauft1} is bounded from above by a constant which is
independent of $ L \in \mathbb{N}$.
\end{proof}

\begin{lemma}\label{lemma:maxdel}
Under assumption~{\bf A1} for any $ s \in (0,1/4) $
\begin{equation}
      \sup_{ z \in \mathbb{C}^+} \sup_{L \in \mathbb{N}}\;
     \mathbb{E}\left[ \frac{1}{|\mathcal{T}_L|^s} \left|
\big\langle \delta_0 , \big| H_{\mathcal{T}_L} - z \big|^{-2}
\delta_0 \big\rangle\right|^s \right] < \infty \, .
\end{equation}
\end{lemma}
\begin{proof}
A combination of \eqref{eq:3folgt1}, \eqref{eq:remainder} (with $ w =
1 $) and \eqref{eq:1folgt3} below yields for all $ z \in
\mathbb{C}^+$ and $ L \in \mathbb{N} $
\begin{multline}
      |\mathcal{T}_L|^{-1} \big\langle \delta_0 , \big|
H_{\mathcal{T}_L} - z \big|^{-2} \delta_0 \big\rangle
      \leq \Im \big\langle \delta_0 , \big( H_{\mathcal{T}_L} -
\Re z -i \, |\mathcal{T}_L|^{-1}\big)^{-1} \delta_0 \big\rangle \\
     \times
     \left( 2 + |\mathcal{T}_L|^{-2} \dist\left( \sigma(
H_{\mathcal{T}_L}), z \right)^{-2} \right) \, .
\end{multline}
We now take the fractional-moment and apply the Cauchy-Schwarz
inequality. The claim then follows from the fractional-moment bound
\eqref{eq:smombound}
and Wegner's estimate \eqref{eq:Wegner}.
\end{proof}

\begin{proof}[Proof of Theorem~\ref{thm:loc}]
We pick an arbitrary bounded Borel set $ I \subset \mathbb{R} $. By
the strong resolvent convergence,
\begin{equation}
      \lim_{L \to \infty} \big\| \big( H_{\mathcal{T}_L} -z
\big)^{-1} \delta_x  - \big( H_{\mathcal{C}} -z \big)^{-1} \delta_x
\big\| = 0
\end{equation}
for all $ x \in \mathcal{C} $ and all $ z \in \mathbb{C}^+ $, and
monotone convergence, it follows from \eqref{eq:finits} that for
Lebesgue-almost all $ E \in I $
\begin{equation}
     \mathbb{E}\left[ \left(\lim_{\eta \downarrow 0} \left\| \big(
H_{\mathcal{C}} - E -i \eta \big)^{-1} \delta_x \right\|^2\right)^{s}
\right]
      < \infty \, ,
\end{equation}
with the same $ s $ as in \eqref{eq:finits}. Since the conditional
distribution of $ \omega_x $ -- conditioned on the sigma-algebra
generated by
$ \{ \omega_y \}_{y\neq x} $ -- has a bounded density, $ \varrho $,
the Simon-Wolff localization criterion \cite[Thm.~8]{SiWo86} is thus
satisfied and yields the assertion.
\end{proof}    

\section{The spectra of random operators on single-ended trees}

On the canopy tree $ \mathcal{C} $ from any site there is a unique
path to infinity, i.e., in the terminology of \cite{Woe00} it is a
single-ended tree.  The purpose of this section is to clarify that
this point in itself implies part of above  localization statement,
but not all of it.
To place Theorem~\ref{thm:loc} in a more general context
we  prove here that  on single-ended trees random operators of the
kind considered here have no
absolutely continuous spectrum, but singular continuous spectrum can
occur (though not in the specific case of the canopy graph).\\

\subsection{Absence of absolutely continuous spectrum}

The main result of the present subsection is proven for the  class of
graphs defined next.

\begin{definition}\label{def:backbonegraph}
For a graph $\mathcal{ G}$ a \emph{backbone} $ \mathcal{B} $ is a 
connected path, indexed by either
$\Z$, $\N$, or $\{1,...,L\}$, whose deletion  transforms $\mathcal{
G}$ into a collection of finite disconnected sets.
Graphs with a backbone are referred to as \emph{backbone graphs}.
\end{definition}

Not every graph has a backbone, and in case it does, the backbone is 
not unique (except in the double-ended case) as can
be seen by considering the canopy graph.
We shall consider below
self-adjoint random operators
\begin{equation}\label{defHB}
H_\mathcal{G,B} = A + W + V
\end{equation}
acting on the Hilbert space $ \ell^2(\mathcal{G}) $ over a backbone graph,
where $ A $ denotes the adjacency operator and $ W $ stands for an
arbitrary multiplication operator. Moreover,  $ V $ denotes the
random multiplication operator which acts only along the backbone
with values at sites $x\in \mathcal{B} $ given by random variables $ 
\{ \omega_x
\}_{x \in \mathcal{B}} $ which we assume to be independent and
identically distributed, with a distribution
satisfying the following assumption.
\begin{enumerate}
\item[] {\bf Assumption A1':}~~ The distribution of the variables
$\omega_x$  is of bounded density, $ \varrho \in L^\infty(\mathbb{R})
$, and satisfies
      $ \int_\mathbb{R} (1+ | \omega_0 |)^2 \varrho(\omega_0)^2 d
\omega_0 < \infty $.
\end{enumerate}
Note that Assumption~{\bf A1} with $ \tau = 2 $ implies Assumption~{\bf A1'}.

\begin{theorem}[Absence of ac spectrum]  \label{thm:noac}
Assuming~{\bf A1'}, for any operator $ H_\mathcal{G,B} = A + W + V $
on the Hilbert-space $ \ell^2(\mathcal{G}) $ associated with a
backbone graph the absolutely continuous component of the spectrum is
empty.
\end{theorem}

In the proof  we shall make use of the following extension of a bound
which was proven for one dimensional random operators in the work of
Delyon, Kunz, and Souillard~\cite{DeKuSo83}.  For this result we
assume the structure described above, except that it is not necessary
for the backbone to extend to infinity.

\begin{proposition}[Exponential decay II]\label{prop:DKS}
On a finite graph $ \mathcal{G}$ with a backbone  $ \mathcal{B}$, let
$ H_\mathcal{G,B} = A + W + V $ be a random operator   of the form
described above.  Assuming {\bf A1'} holds, for any bounded interval
$ I \subset \mathbb{R} $ and $ s \in (0,1/2) $ there exists $ C(s,I)
< \infty $ such that for any  pair of sites along the backbone
$x,y\in \mathcal{B} $:
\begin{equation}\label{eq:DKS}
\int_I \mathbb{E}\left[ \left| \langle \delta_x , \left(
H_{\mathcal{G,B}} - E \right)^{-1} \delta_y \rangle \right|^s 
\right] dE
\leq C(s,I) \, \exp\left( - s \, \lambda \, {\rm dist}(x,y) \right)  
\end{equation}
where (as in  \cite[Eq.~(1.8)]{DeKuSo83})
\begin{equation}\label{def:lambda}
\lambda := \inf_{\substack{\eta > 0 \\ 40 \eta |\log \eta | <
1}} -2 \big/ \log\left(1 - \frac{\alpha(\eta)}{25}(1 - 40 \eta |\log
\eta |)^2 \right) \, ,
\end{equation}
with $ \alpha(\eta) := 1 - \sup_{|\xi|>\eta} | \hat \varrho(\xi)| $,
and $ \hat \varrho $ the Fourier transform of the single-site density.
\end{proposition}

Although it is not stated there in the above form, this exponential
bound readily follows from the one-dimensional analysis of
\cite{DeKuSo83}.    We present this reduction in
Appendix~\ref{app:DKS}.   Assuming  Proposition~\ref{prop:DKS}  we
turn now to the derivation of the main result of this section.

\begin{proof}[Proof of Theorem~\ref{thm:noac}]
We shall give the proof for $ \mathcal{B} $ single-ended; the 
case of a double-ended backbone follows similarly and,
in case $  \mathcal{B} $ is finite there is nothing to show.   

Let $ x_0 \in \mathcal{B} $ be arbitrary and pick $ x_L \in 
\mathcal{B} $ in the direction towards infinity with $ {\rm 
dist}(x_0,x_L) = L $.
By removing the bond between $ x_L $ and $ x_{L+1} $, we cut $ 
\mathcal{G} $ and its backbone into
finite parts $ \mathcal{G}_L  $, $ \mathcal{B}_L $ and the infinite remainders.
Let $ H_{\mathcal{G}_L,\mathcal{B}_L} $ denote the restriction of $ 
H_\mathcal{G,B} $ to $ \ell^2(\mathcal{G}_L) $.

The resolvent identity and the fact that
$ \Im \,  \langle \delta_{x_0} , (H_{\mathcal{G}_L,\mathcal{B}_L} - E 
)^{-1} \delta_{x_0}\rangle = 0 $ for almost all $ E \in \mathbb{R} $ 
implies
\begin{align}
	\Im \, \langle \delta_{x_0} , (H_\mathcal{G,B} - E - i 0 
)^{-1} \delta_{x_0} \rangle
	 \leq &
	\left| \langle \delta_{x_0} ,  (H_{\mathcal{G}_L, 
\mathcal{B}_L} - E )^{-1} \delta_{x_L} \rangle \right|^2 \notag \\
	& \times \left| \langle \delta_{x_{L+1}} , (H_\mathcal{G,B} - 
E - i 0 )^{-1} \delta_{x_{L+1}}\rangle  \right| \label{eq:proofac0}
\end{align}
Since the ac component of the spectral measure of $ H_\mathcal{G,B} $ 
associated with $ \delta_{x_0}$ is supported on those $ E \in 
\mathbb{R} $ for which the
left-hand side is finite and strictly positive, it remains to show 
that the right-hand side converges as $ L \to \infty $ to zero in 
distribution with respect to
the product of the probability measure associated with $ \{ \omega_x 
\} $ and Lebesgue measure for $ E \in I $, where $ I \subset 
\mathbb{R} $ is an arbitrary  bounded interval.
For a proof of the latter we note that the second term on the 
right-hand side of \eqref{eq:proofac0} is seen to be bounded in 
probability using,
for example, a fractional moment estimate, cf. \eqref{eq:smombound}.
Moreover, Proposition~\ref{prop:DKS} implies that the first term 
converges to zero in distribution with respect to the above product 
measure.
\end{proof}

\subsection{Appearance of singular continuous spectrum}

\begin{figure}[hbt]
\hfill\begin{minipage}{.4\textwidth}
\begin{picture}(0,0)%
\includegraphics{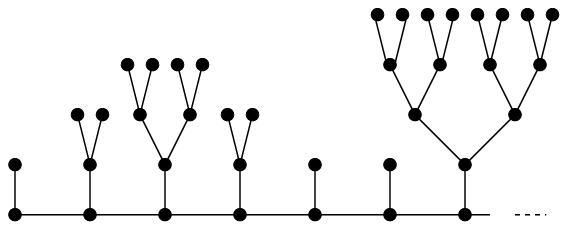}%
\end{picture}%
\setlength{\unitlength}{1579sp}%
\begingroup\makeatletter\ifx\SetFigFont\undefined%
\gdef\SetFigFont#1#2#3#4#5{%
   \reset@font\fontsize{#1}{#2pt}%
   \fontfamily{#3}\fontseries{#4}\fontshape{#5}%
   \selectfont}%
\fi\endgroup%
\begin{picture}(6833,3089)(601,-5467)
\put(1501,-5386){\makebox(0,0)[lb]{\smash{{\SetFigFont{7}{8.4}{\rmdefault}{\mddefault}{\updefault}{\color[rgb]{0,0,0}$x_1$}%
}}}}
\put(601,-5386){\makebox(0,0)[lb]{\smash{{\SetFigFont{7}{8.4}{\rmdefault}{\mddefault}{\updefault}{\color[rgb]{0,0,0}$x_0$}%
}}}}
\put(2476,-5386){\makebox(0,0)[lb]{\smash{{\SetFigFont{7}{8.4}{\rmdefault}{\mddefault}{\updefault}{\color[rgb]{0,0,0}$x_2$}%
}}}}
\put(3376,-5386){\makebox(0,0)[lb]{\smash{{\SetFigFont{7}{8.4}{\rmdefault}{\mddefault}{\updefault}{\color[rgb]{0,0,0}$x_3$}%
}}}}
\put(4201,-5386){\makebox(0,0)[lb]{\smash{{\SetFigFont{7}{8.4}{\rmdefault}{\mddefault}{\updefault}{\color[rgb]{0,0,0}$x_4$}%
}}}}
\put(6076,-5386){\makebox(0,0)[lb]{\smash{{\SetFigFont{7}{8.4}{\rmdefault}{\mddefault}{\updefault}{\color[rgb]{0,0,0}$x_6$}%
}}}}
\put(5101,-5386){\makebox(0,0)[lb]{\smash{{\SetFigFont{7}{8.4}{\rmdefault}{\mddefault}{\updefault}{\color[rgb]{0,0,0}$x_5$}%
}}}}
\end{picture}%
\end{minipage}\hfill
\begin{minipage}{.3\textwidth}
\caption{Example of the construction of a single-ended tree 
$  \mathcal{G} $ obtained by 
 gluing finite trees to the sites of a backbone.}\label{fig:backbone}
\end{minipage}
\end{figure}

We shall now focus on backbone graphs which are obtained by decorating 
an infinite path $ \mathcal{B} $ with trees, as  
in  Figure~\ref{fig:backbone}. The canopy graph is within this class but 
our main result applies to graphs where the trees attached to the line 
grow at a  much faster rate.    
The goal is to prove that there are single-ended trees 
$ \mathcal{G} $ for which random operators of the form
\begin{equation}
H_\mathcal{G,B} = A + V
\end{equation}
have a singular continuous component in their spectrum. Here and 
in the following, we will assume that $ V $ is
the multiplication operator corresponding to independent and 
identically distributed random variables $ \{ \omega_x \}_{x \in 
\mathcal{G} } $
on the \emph{whole} tree.

\begin{theorem}[Singular continuous spectrum]\label{thm:singular}
Fix a bounded, open interval
   $  I \subset \, \Sigma_{\rm cont}(H_\mathcal{T}) $ where $ 
H_\mathcal{T} = A + V $ with random variables satisfying {\bf A1}.
   Then there exist a single-ended tree $ \mathcal{G} $ for which $ 
H_\mathcal{G,B} = A + V $ has singular
continuous spectrum on $ I $.
\end{theorem}
Two remarks apply:
\begin{enumerate}
\item  The existence of continuous spectrum, $ \Sigma_{\rm 
cont}(H_\mathcal{T})  \neq \emptyset $
for the random operator $ H_\mathcal{T} = A + V $ on the infinite 
rooted regular tree $ \mathcal{T} $ for small disorder is ensured by 
\cite{Kle95,Kle98,ASW05,FrHaSp06}.
\item It has been noted before that random operators with decaying 
potential \cite{Del85} or the Laplacian on certain (tree) graphs 
\cite{Si96,Bre06} may exhibit singular continuous spectrum.
\end{enumerate}

The construction of trees $ \mathcal{G} $ in 
Theorem~\ref{thm:singular} is based on the
following observation.
\begin{lemma}\label{lemma:explode}
Assume {\bf A1} and let $ I \subset \, \Sigma_{\rm 
cont}(H_\mathcal{T}) $ be a bounded Borel set and $ s \in (0, \tau/2] 
$. Then
\begin{equation}
	\lim_{L\to \infty}  \; \int_I \mathbb{E} \left[ \langle \delta_0 , ( 
H_{\mathcal{T}_L} -E)^{-2} \delta_0 \rangle^{-s} \right] dE   = 0 \, .
\end{equation}
\end{lemma}
\begin{proof}
We pick $ \eta > 0 $. The spectral theorem implies that for every $ L 
\in \mathbb{N} $
\begin{align}
\langle \delta_0 , ( H_{\mathcal{T}_L} -E)^{-2} \delta_0 \rangle^{-1} & \leq
  \langle \delta_0 , \left[ ( H_{\mathcal{T}_L} -E)^{2} + \eta^2 
\right]^{-1} \delta_0 \rangle^{-1} \\
  & \leq \langle \delta_0 ( H_{\mathcal{T}_L} -E)^{2} \delta_0 \rangle + \eta^2
  	\leq 4 K + \omega_0^2 + E^2 + \eta^2 \, . \notag
\end{align}
Using the dominated convergence theorem and strong resolvent 
convergence we thus conclude
\begin{align}
	\lim_{L\to \infty}  \; & \int_I \mathbb{E} \left[ \langle 
\delta_0 , ( H_{\mathcal{T}_L} -E)^{-2} \delta_0 \rangle^{-s} \right] 
dE \notag \\
		& \leq \int_I \mathbb{E} \left[ \langle \delta_0 , 
\left[ ( H_{\mathcal{T}} -E)^{2} + \eta^2 \right]^{-1} \delta_0 
\rangle^{-s} \right] dE \, .
\end{align}
The claim follows with the help of the the dominated convergence 
theorem from the fact that for Lebesgue almost all $ E \in 
\Sigma_{\rm cont}(H_\mathcal{T}) $
\begin{equation}
\lim_{\eta \downarrow 0} \langle \delta_0 , \left[ ( H_{\mathcal{T}} 
-E)^{2} + \eta^2 \right]^{-1} \delta_0 \rangle^{-1} = 0 \, .
\end{equation}
\end{proof}

The proof of Theorem~\ref{thm:singular} is also based on the 
following  Simon-Wolff type criterion, which may be deduced from 
\cite{RSS94}.

\begin{proposition}[cf.~\cite{RSS94}]\label{prop:SW}
  Let $ H $ be a random operator on the Hilbert space over a graph, 
with iid random potential whose single site distribution is ac and 
bounded.
Assume  there is  a Borel set $ I $ such that
\begin{enumerate}
\item H has no absolutely continuous spectrum in $ I $.
\item  for Lebesgue almost every $ E \in I $ almost surely the Green
function is not square summable:
\begin{equation}\label{eq:SW}
\lim_{\eta \downarrow 0}  \, \langle \delta_{x} , [(H - E)^2 + \eta^2 
]^{-1} \delta_{x} \rangle^{-1} = 0 \, .
\end{equation}
\end{enumerate}
Then, in the space for which $ \delta_{x} $ is a cyclic vector, 
almost surely $ H $ has only singular continuous spectrum in $ I $.
\end{proposition}

Finally, our construction is also based on the following lower bound 
on the decay of  the Green function of single-ended graphs $ 
\mathcal{G} $
along the backbone $ \mathcal{B} $. It is important for us that the 
decay rate is controlled independently of the depth of the trees 
glued to $  \mathcal{B}   $.
\begin{lemma}\label{lemma:lowerdecay}
Assume {\bf A1} and let $ E_0 > 0 $ and $ s \in (0,\min\{\tau,1/2\}] 
$. There exist some constant $ C(s,E_0) < \infty $ such that
for all $ x_0 , x \in \mathcal{B} $
  \begin{align}
	\sup_{|E|\leq E_0 } \sup_{\eta \in (0,1)} & \mathbb{E}\left[
		\left| \langle \delta_{x_0} , \left( H_\mathcal{G,B} 
- E - i \eta \right) \delta_x \rangle \right|^{-s} \, 	\big|  \,
	\{ \omega_x \}_{x \not\in \{ y : {\rm 
dist}(y,\mathcal{B})\leq 1 \}} \right]  \notag \\
	& \leq C(s,E_0) \; e^{\lambda(s,E_0) \, {\rm dist}(x_0,x) } \, .
\end{align}
Here
\begin{equation}\label{def:lambda2}
\lambda(s,E_0) := 1 + \log \left( 1 + E_0 + \int |\omega_0|^s 
\varrho(\omega_0) d\omega_0 + K' C_s \right)  \, ,
\end{equation}
where $ K' $ is the maximal number of vertices neighboring $ 
\mathcal{B} $, and $ C_s $ is the constant appearing in 
\eqref{eq:smombound}.
\end{lemma}
\begin{proof}
Similarly as in \eqref{eq:Gfactor} (cf.\ \cite{Kle98}) we factorize 
the Green function into a product,
\begin{align}\label{eq:factor}
     & \big\langle \delta_{x_0} , \big( H_{\mathcal{G,B}} - E - i \eta 
\big)^{-1} \delta_x \big\rangle = \prod_{j=0}^{\dist(x_0,x)}
\Gamma_{j} \\
     & \qquad \mbox{with} \quad \Gamma_{j} := \big\langle
\delta_{x_j} , \big( H_{\mathcal{G}(x_j)} - E \big)^{-1}
\delta_{x_j} \big\rangle \, . \notag
\end{align}
Here $ x_0, x_1 , \dots , x_{ \dist(0,x) } := x $ are the
vertices on $ \mathcal{B} $ connecting $ x_0 $ with $ x $.
Moreover,
$ \mathcal{G}(x_j) $ is that infinite subtree of $ \mathcal{G} $ 
which is forward to $ x_j $ in the direction away from $ x_0 $.
The factors in \eqref{eq:factor} satisfy the following relation (cf.\ 
\cite{Kle98})
\begin{align}
  \Gamma_{j}^{-1} = & V_{x_j} - E - i \eta - \Gamma_{j+1} - G_j \\
  	& \mbox{where}\quad
	G_j  := \sum_{\substack{{\rm dist}(w,x_j)= 1 \\ w \not\in 
\mathcal{B}}} \big\langle
\delta_{w} , \big( H_{\mathcal{G}(w)} - E \big)^{-1}
\delta_{w} \big\rangle \, , \notag
\end{align}
and the sum is over all neighboring vertices of $ x_j $ which are not 
on $ \mathcal{B} $ and each term involves the finite subtree tree $ 
\mathcal{G}(w) $ which is rooted at
$ w $ and extends away from the backbone.

We now integrate the product in \eqref{eq:factor} step by step 
starting with estimating the conditional
expectation, conditioning on all random variables aside from $ x_0 $ 
and its neighbors which do not belong to $ \mathcal{P}(x_0) $,
\begin{align}
	&\mathbb{E}\left[ \left|\Gamma_{0}\right|^{-s}  \big|
		\{\omega_{x_0},\omega_w\}_{\{ w \not\in \mathcal{B} 
:   {\rm dist}(w,x_0)= 1 \} }^c \right] \notag \\
	& \leq 	\int |\omega_0|^s \varrho(\omega_0) d\omega_0 + |E| + 
\eta + |\Gamma_1|^{-s}
		+ K' C_s \notag \\
	& \leq \exp( \lambda(s,E_0) - 1 ) + |\Gamma_1|^{-s} \, .
\end{align}
Here the last factor in the second line stems from estimating the 
expectation of the terms contributing to $ G_0 $ using 
\eqref{eq:smombound}.
Iterating this bound yields the result.
\end{proof}

\begin{proof}[Proof of Theorem~\ref{thm:singular}]
Let $ E_I = \max\{ |E| \, : \, E \in I \} $ and $ \tau' := 
\min\{\tau,1/2\}/2 $.
We define a sequence of finite regular trees $ \mathcal{T}_{L_n} $
through the requirement
\begin{equation}\label{def:Ln}
	\int_I \mathbb{E} \left[ \langle \delta_{0} , ( 
H_{\mathcal{T}_{L_n}} -E)^{-2} \delta_{0} \rangle^{-\tau'} \right] dE 
\leq \exp\left( - 2 \lambda(2 \tau',E_I) \, n \right)
\end{equation}
for every $ n \in \mathbb{N}_0 $, where $ \lambda(\tau'/4,E_I) $ is 
the constant appearing in \eqref{def:lambda2}. Note that such a 
sequence exists thanks to Lemma~\ref{lemma:explode}.\\

Given a half-infinite line $ \mathcal{B} $ we glue to every vertex $ 
x_n \in  \mathcal{B} $ with $ {\rm dist}(x_0,x_n) = n $ another edge 
which connects
$ x_n $ with the root of the tree $ \mathcal{T}_{L_n} $; cf.\ 
Figure~\ref{fig:backbone}.\\

To conclude that the spectral measure of $ H_{\mathcal{G,B}} $ 
associated with $ \delta_{x_0} $ is purely
singular continuous in $ I $, we use Proposition~\ref{prop:SW} and 
first note that for any $ L \in \mathbb{N} $ and Lebesgue-almost 
every $ E \in \mathbb{R} $
\begin{align}
	& \liminf_{\eta \downarrow 0} \,
		\langle \delta_{x_0} ,  [(H_{\mathcal{G,B}} - E)^2 + 
\eta^2 ]^{-1}  \delta_{x_0} \rangle \notag \\
	&  \geq \liminf_{\eta \downarrow 0}     \sum_{x \in \mathcal{B}_L}
	\left|  \langle \delta_{x_0}  ,
	\left( H_{\mathcal{G,B}} - E - i \eta \right)^{-1} 
\delta_{x} \rangle \right|^2  \notag \\
	& = \sum_{n=0}^L
	\left|  \langle \delta_{x_0} ,
		 \left( H_{\mathcal{G,B}} - E - i 0 \right)^{-1} 
\delta_{x_n} \rangle \right|^2
		 \left( 1 +  	S(n)   \right) \\
		 & \qquad\qquad \mbox{where} \quad
		 S(n) := \langle \delta_0 , ( H_{\mathcal{T}_{L_n}} 
-E)^{-2} \delta_0 \rangle \, . \notag
\end{align}
We now average an inverse power of the above quantity and integrate over $ I $.
Jensen's inequality thus yields
\begin{align}
	& \int_I \mathbb{E}\left[  \limsup_{\eta \downarrow 0} \,
		\langle \delta_{x_0} ,  [(H_{\mathcal{G,B}} - E)^2 + 
\eta^2 ]^{-1}  \delta_{x_0} \rangle^{-\tau'/2}
			\right] dE \notag \\
	& \leq  \frac{1}{L^{\tau'+1}} \sum_{n=0}^L \int_I 
\mathbb{E}\left[  \left|  \langle \delta_{x_0} ,
		 \left( H_{\mathcal{G,B}} - E - i 0 \right)^{-1} 
\delta_{x_n} \rangle \right|^{-\tau'}
		  S(n)^{-\tau'/2} \right] dE  \, . \label{eq:scproof}
\end{align}
The Cauchy-Schwarz inequality together with \eqref{def:Ln} and 
Lemma~\ref{lemma:lowerdecay} imply that the right-hand side in 
\eqref{eq:scproof} is bounded from above by
\begin{equation}
	\frac{1}{L^{\tau'+1}}  \, C(2\tau',E_0)^{1/2} \, \sum_{n=0}^L
			\exp\left( - \lambda(2\tau',E_0) \, 
\frac{n}{2} \right) \, ,
\end{equation}
which converges to zero as $ L \to \infty $. This yields the claimed result.
\end{proof}

The lengths of the regular trees $ \mathcal{T}_{L_n} $ are defined 
via \eqref{def:Ln}
in a rather indirect way. In particular, no estimates are given. To 
answer the question about the minimal growth of $L_n $ as $ n \to 
\infty $, which is sufficient for the
production of singular continuous spectrum, one needs to estimate the 
growth of the quantity
$ \langle \delta_{0} , ( H_{\mathcal{T}_{L}} -E)^{-2} \delta_{0} 
\rangle $ as $ L \to \infty $ for
$ E \in \Sigma_{\rm cont}(H_\mathcal{T} )$.

\vskip 2cm

\appendix

\noindent {\Large \bf  Appendix} \mbox{ } \\

For completeness, in the following appendix sections  we shall 
briefly sketch proofs of some spectral properties of the canopy graph which 
are of relevance to our discussion and which are derived by arguments 
which are already in the literature.   We also add some observations 
and discussion.

\section{Some basic properties of the canopy operator}

\subsection{Existence of the canopy density of states measure}\label{App:Erg2}

Following is a brief sketch of the proof of Theorem~\ref{thm:DOS}.

\begin{proof}[Proof of Theorem~\ref{thm:DOS}]
We embed $ \mathcal{T}_L $ into $ \mathcal{C} $ so that $ \partial 
\mathcal{T}_L \subset \partial \mathcal{C} $.
The trace in \eqref{eq:dos} can be decomposed
into contributions from layers with a fixed distance to the outer boundary,
\begin{align}
     |\mathcal{T}_L |^{-1} \Tr F(H_{\mathcal{T}_L})
      & = \frac{K-1}{K} \sum_{n=0}^{L} K^{-n}  \; T_{n,L}(F) \\
     & \mbox{where} \quad T_{n,L}(F)  :=
     K^{n+1-L} \mkern-20mu \sum_{x \, : \,
\dist(x,\partial\mathcal{T}_L)=n} \mkern-10mu
     \langle \delta_x \, ,  F(H_{\mathcal{T}_L}) \, \delta_x
\rangle \, . \notag
\end{align}
Each contribution $ T_{n,L}(F) $
is normalized to one for $ F = 1 $ and, more generally, $  T_{n,L}(F)
\leq \| F \|_\infty $.
Thanks to dominated convergence, it is therefore enough to prove the
following almost-sure convergence
for each $ n \in \mathbb{N}_0 $
\be\label{eq:layer}
\lim_{L \to \infty} T_{n,L}(F)
     = \mathbb{E}\left[ \langle \delta_{x_n} , F(H_\mathcal{C}) \,
\delta_{x_n} \rangle \right] \, ,
\ee
where $ x_n \in \mathcal{C} $ is an arbitrary vertex with $ \dist(x_n
,\partial\mathcal{C})=n $, cf.\ Figure~\ref{Fig:Canopy}.\\

The proof of~\eqref{eq:layer} boils down to the Birkhoff-Khintchin
ergodic theorem \cite{Kal02}
and an approximation argument.
Since the functions $ \varphi_z = ( \cdot - z )^{-1} $ with $ z \in
\mathbb{C}^+ $  are dense in $ C_b(\mathbb{R}) $ and
the linear functionals in both sides of \eqref{eq:layer} are
(uniformly) continuous on $ C_b(\mathbb{R}) $,
it is sufficient to prove \eqref{eq:layer} for $ F = \varphi_z $.

By truncating $ \mathcal{T}_L $ at a layer $ n + L_0 $ below the
outer boundary, we may approximate the sum $ T_{n,L}(\varphi_z) $ by
$ K^{L-n-L_0} $
stochastically independent terms of the form
\be\label{eq:approxterm}
     \frac{1}{K^{L_0}} \sum_{\dist(0,y) = L_0} \langle \delta_y \,
, \varphi_z(H_{\mathcal{T}_{n+L_0}}) \, \delta_y \rangle \, .
\ee
The approximation error can be kept arbitrarily small by taking $ L_0
\in \mathbb{N} $ large.
The approximating average of $ K^{L-n-L_0} $  stochastically
independent terms
satisfies the assumptions of the
Birkhoff-Khintchin ergodic theorem for iid random variables. As $ L
\to \infty $, it therefore
converges almost surely to
\be
      \mathbb{E}\left[     \frac{1}{K^{L_0}} \sum_{\dist(0,y) =
L_0} \langle \delta_y \, , \varphi_z(H_{\mathcal{T}_{n+L_0}}) \,
\delta_y \rangle \right]
     = \mathbb{E}\left[ \langle \delta_{x_n} ,
\varphi_z(H_{\mathcal{T}_{n+L_0}}) \, \delta_{x_n} \rangle \right] \,
,
\ee
where $ x_n $ is an arbitrary vertex in the $ n$th layer below the
surface $ \partial\mathcal{T}_{n+L_0}$ .
Taking $ L_0 \to \infty $, the last term converges to the right side
in~\eqref{eq:layer} by the dominated convergence theorem.
\end{proof}

The standard Wegner estimate  allows to conclude some regularity of $ n_\mathcal{C} $.
\begin{lemma}\label{lemma:dos}
Under the assumption~{\bf A1} the canopy density of states 
measure $ n_\mathcal{C} $ is
absolutely continuous, with bounded density satisfying  
     \be
         d_\mathcal{C}(E) = \frac{n_{\mathcal{C}}(dE)}{dE}
\leq \| \varrho \|_\infty \, .
     \ee
\end{lemma}
\begin{proof}
This readily follows from the definition of the density of states  measure  \eqref{eq:dos} and the Wegner estimate \eqref{eq:Wegner}.
\end{proof}

\subsection{Spectrum of the adjacency operator on the canopy 
graph}\label{App:Erg}
We will now give a brief sketch of the proof of the following assertion:

\begin{proposition}[cf.\ \cite{AF00}]
The spectrum of of the adjacency operator with boundary
condition $ A + B $ on $ \ell^2(\mathcal{C}) $
consists of infinitely degenerate eigenvalues coinciding with the
union of all
eigenvalues of the adjacency operator (with constant
boundary condition $ b \in \mathbb{R} $) on $ \ell^2(\{1,2,\dots,n\}) $
with $ n \in \mathbb{N} $ arbitrary. The corresponding eigenfunctions
are compactly supported.
\end{proposition}

\begin{proof}
The basic construction is simplest to describe for $K=2$.    In that case, 
for each eigenfunction $\psi$ of the adjacency operator on $ \ell^2(\{1,2,\dots,n\}) $ 
eigenfunctions can be constructed on the canopy graph which are supported on the forward 
trees 
corresponding to any site $x$ which is at distance $n+1$ from the outer boundary $\partial  \mathcal{C}$.   
The functions are defined so they are antisymmetric with respect to 
the exchange of the two forward trees of length $n$ which lie 
between $x$ and the boundary $\partial  \mathcal{C}$, and on  each of the two 
subtrees are given by a radially 'fanned out' versions of $\psi$ 
(cf.\ \cite[Proof of Prop. A.1]{ASW05}).   It is easy to verify that 
the construction yields a complete orthonormal collection of eigenfunctions.

To determine the spectrum of $ A + B $ on $ \ell^2(\mathcal{C}) $ with a general $K$ 
one may use -- analogously to \cite{AF00} --
  a decomposition of the Hilbert space into invariant subspaces:
\begin{align}\label{eq:Hilbertdec}
     \ell^2(\mathcal{C})  = &\bigoplus_{x \in \mathcal{C}} \,
\mathcal{Q}_x \, , \quad  \mbox{with} \quad
     \mathcal{Q}_x := \Big( \bigoplus_{\substack{y \in
\mathcal{C}(x) \\ \dist(x,y)=1}} \mathcal{S}_y \;\Big) \ominus
\mathcal{S}_x 
\end{align}
where $\mathcal{S}_x$ denotes the subspace of symmetric functions on the forward subtree $
\mathcal{C}(x) $, cf.\ \eqref{eq:subtreedef}: 
\begin{align}
\mathcal{S}_x := \left\{ \psi \in \ell^2\big(\mathcal{C}\big)  : \, 
\begin{array}{l}
         \mbox{$ y \mapsto \langle \delta_y , \psi\rangle \; $ is 
supported on $\mathcal{C}(x) $}\\
         \mbox{and constant on each generation of $\mathcal{C}(x) $} 
\end{array}\right\}  \,.    \notag
\end{align}

The orthogonal decomposition \eqref{eq:Hilbertdec} reduces the
operator $ A + B $ on $ \ell^2(\mathcal{C}) $
to an orthogonal sum of operators on $ \mathcal{Q}_x $, each of which
is unitarily equivalent to the orthogonal sum of $ K-1 $ operators on
$ \mathcal{S}_y $ where $ y $ is one of the forward neighbors  of $ x
$.
In turn, each operator on $ \mathcal{S}_y $ is unitary equivalent to
the adjacency operator (with constant
boundary condition $ b \in \mathbb{R} $) on the Hilbert space $
\ell^2(\{1, 2 , \dots, \dist(y,\partial\mathcal{C}) \}) $.
\end{proof}

As an aside, we note that other examples of discrete operators with 
finitely supported eigenfunctions can be found in \cite{GZ01,DS01}.\\

\section{Negligibility in probability of the spectral measure within
the singular spectrum}\label{app:sing}

It may be of some interest to observe that, as is proven below, the 
condition which by Theorem~\ref{prop:infdiv} implies Poisson 
statistics holds throughout the singular spectrum of the 
infinite-volume operator.   To avoid confusion, let us note that 
although the reference here is to the spectral measure at the root, 
the infinity divisibility which this implies is of the spectral 
measure whose density is given by the canopy dos, weighted as in 
\eqref{dos}.
This observation is not used for our main result since we establish 
the negligibility in probability thorough another mechanism,  which 
is valid throughout the entire spectrum.

The relevant statement is valid not only in the tree
setup and is based on the following 
measure theoretic statement. 

\begin{theorem} \label{lemma:point}
  Suppose the operator $H_\mathcal{T}$ has almost surely  
only singular spectrum in a given Borel set $I$.   Then
  for Lebesgue-almost all $ E \in I $ 
the condition in Definition~\ref{def:unif} is satisfied.
\end{theorem}
\begin{proof}
As $ L \to \infty $ the spectral measure $ \sigma_{x,L} $
converges vaguely  to $ \langle \delta_x ,
P_{\cdot}(H_{\mathcal{T}}) \delta_x \rangle $, which is finite and
purely singular on $ I $.
The subsequent lemma thus implies that $ \mathbb{P} $-almost surely 
for every $ \varepsilon > 0 $ and $ w > 0 $
\begin{equation}\label{eq:assertion}
     \left|  \limsup_{L \to \infty}\left\{ E \in I 
      \, : \,  |\mathcal{T}_L | \, \sigma_{x,L}\big(E
+ |\mathcal{T}_L |^{-1} (-w,w) \big)
       > \varepsilon
         \right\} \right| = 0 \, .
\end{equation}
By  Fubini-Tonelli's theorem this implies that
for
every $ \varepsilon > 0 $ and every $ w > 0 $ there exists a  subset 
$ J(\varepsilon,w) \subset \Sigma_{\rm
sing}(H_\mathcal{T}) $ of full Lebesgue measure such that for all $ 
E \in J(\varepsilon,w) $
\begin{multline}
\lim_{L\to \infty}\mathbb{P} \left( \left\{  |\mathcal{T}_L | \,
\sigma_{x,L}\big(E + |\mathcal{T}_L |^{-1} (-w,w) \big) > \varepsilon
     \right\}\right) \\
\leq \mathbb{P} \left( \limsup_{L\to \infty} \left\{  |\mathcal{T}_L | \,
\sigma_{x,L}\big(E + |\mathcal{T}_L |^{-1} (-w,w) \big) > \varepsilon
     \right\}\right)  = 0  \, .
\end{multline}
Since the event in the right-hand side is monotone in both $ 
\varepsilon $ and $ w $,
we may pick any two monotone sequences $ \epsilon_n \to 0 $ and $ w_m 
\to \infty $ and
define
$ J := \bigcap_{n,m}  J(\varepsilon_n,w_m) $, a set of full Lebesgue 
measure, on which
the claimed convergence \eqref{eq:condition} holds for all $ 
\varepsilon > $ and $ w > 0 $.
\end{proof}

Following is a rather general observation for singular measures.  

\begin{lemma}
Let $ \sigma $ be a purely singular measure on $ I \subset
\mathbb{R} $, suppose that $ \lim_{n \to \infty} \sigma_n = \sigma $
vaguely, and let
$ \left\{ \xi_n \right\}_{n=0}^\infty $ be a null sequence.
Then for every $ \varepsilon > 0 $ and $ w > 0 $ the sequence 
of sets 
$$  A_n(\varepsilon,w) := \left\{ E \in I \, : \,
\sigma_n\left(E-w \, \xi_n, E+w \, \xi_n\right) > \varepsilon \,
\xi_n \right\} $$  
satisfies 
\begin{equation}\label{eq:asser}
     \left| \limsup_{n \to \infty} A_n(\varepsilon,w) \right| =
\left| \bigcup_{n = 0}^\infty \bigcap_{m = n}^\infty
A_n(\varepsilon,w) \right| = 0 \, . 
\end{equation}
\end{lemma}
\begin{proof} We prove the assertion by contradiction. Suppose there
exists $ \varepsilon > 0 $, $ w > 0 $,  $ M \in \mathbb{N} $ such that
\begin{equation}
        \left| \bigcap_{m = M }^\infty A_n(\varepsilon,w) \right| > 0 \, .
\end{equation}
This implies that there exists an open ball $ B \subset \bigcap_{m =
M }^\infty A_n(\varepsilon,w) $. By assumption $ \sigma $ is purely
singular on this ball, such
that for every $ \delta > 0 $ there exists a finite collection of
disjoint closed intervals $ \{ I_k^\delta \}_{k= 1}^{N_\delta} $,
each of which is contained in $ B $, such that \cite{Kal02}
\begin{equation}
     \left|B \setminus  \bigcup_{k= 1}^{    N_\delta} I_k^\delta
\right| < \delta \quad\mbox{and}\quad
     \sigma\left( \bigcup_{k= 1}^{    N_\delta} I_k^\delta \right)
< \delta  \, .
\end{equation}
Since the above intervals are closed, vague convergence implies  
\begin{equation}\label{eq:vague}
     \sigma\left( \bigcup_{k= 1}^{N_\delta} I_k^\delta \right)
\geq \limsup_{n \to \infty} \, \sigma_n\left( \bigcup_{k= 1}^{
     N_\delta} I_k^\delta \right)
     = \sum_{k= 1}^{    N_\delta}  \limsup_{n \to \infty} \,
\sigma_n\left( I_k^\delta \right) \, .
\end{equation}
Since the intervals are contained in $ \bigcap_{m = M }^\infty
A_n(\varepsilon,w) $, it follows by a covering argument
that for every $ \delta > 0 $ and $ k \in \{ 1, \dots , N_\delta \} $
\begin{equation}
     \limsup_{n \to \infty} \, \sigma_n\left( I_k^\delta \right)
\geq \frac{\varepsilon}{2} \left|  I_k^\delta \right| \, .
\end{equation}
Inserting this inequality in \eqref{eq:vague}, we thus obtain $
\delta > \frac{\varepsilon}{2} \left(\left| B \right| - \delta
\right) $,
which yields a contradiction for $ \delta $ small enough.
\end{proof}



\section{A decorated Delyon-Kunz-Souillard bound}\label{app:DKS'}

It is rather generally true that eigenfunctions of one-dimensional random operators 
decay exponentially.   In particular, 
Delyon, Kunz and Souillard \cite{DeKuSo83} (see also \cite{KS80}) presented a proof of localization 
which does not require  translational covariance.  
Their result can be used in a fairly  straightforward way to imply  Proposition~\ref{prop:DKS}, 
which asserts a  generalization of their result to  
exponential decay of the Green function  on graphs which arise from an arbitrary decorations of 
 a line with finite graphs.  
 The statement which is more directly related to the result of \cite{KS80} is the exponential decay of the eigenfunction correlation.   We shall first prove that and then use it to derive Proposition~\ref{prop:DKS}.

\begin{proposition}\label{prop:DKS2}
In the setting of Proposition~\ref{prop:DKS} there exists $ C
< \infty $ such that for any bounded interval
$ I \subset \mathbb{R} $ and any 
$x,y\in \mathcal{B} $:
\begin{equation}\label{eq:DKS2}
\mathbb{E}\left[ \sum_{E \in I \cap {\rm spec} H_{\mathcal{G}, \mathcal{B} }}  \left|\psi_E(x)\right| \left|\psi_E(y)\right|\right] dE
\leq C \, |I | \, \exp\left( -  \lambda \, {\rm dist}(x,y) \right)  
\end{equation}
with $ \lambda > 0 $ as in \eqref{def:lambda}, which is independent of $ I $. 
\end{proposition}

\begin{proof}  
We proceed by relating Proposition~\ref{prop:DKS2} with a result in \cite{DeKuSo83}.   
In essence, the point is that for a graph $\mathcal{G}$ with random potential along the 
backbone $\mathcal{B}$, the restriction of an eigenfunction to $\mathcal{B}$ coincides 
with an eigenfunction of a one dimensional operator for which the rest of $\mathcal{G}$ 
provides an energy dependent potential.  The analysis of \cite{DeKuSo83} is carried 
pointwise in energy, and hence it is applicable also to the backbone graphs which are 
considered here. 

Let us start by recalling  a change of variable formula, for which   
$\mathcal{B} $ can be any finite subgraph of a graph $ \mathcal{G} $.  
Let $ T $ be an arbitrary self-adjoint operator in $ \ell^2(\mathcal{G}) $ and 
suppose that  $ V $ a random multiplication operator whose 
potential variables $ \{\omega_x \}_{x \in \mathcal{B} } $ are independently distributed with densities $   \varrho_x \in L^1(\mathbb{R}) $.   
Moreover, let    $ \psi_E  $ be  the $ \ell^2(\mathcal{G}) $-normalized eigenfunctions of  
$ H_\mathcal{G,B} = T + V $ at eigenvalues $ E \in {\rm spec} H_\mathcal{G,B} $.  The 
the following relation holds between probability averages:
\begin{multline}  \label{eqnasty}
  \int \prod_{x \in \mathcal{B}} \varrho_x(\omega_x) d\omega_x   
  \sum_{E \in {\rm spec}( H_\mathcal{G,B})}   \left| \psi_E(x_0) \right|^2 \,   
  (\cdots  )
  \\ 
   = \int   \prod_{x \neq x_0} \varrho_x(\omega_x) d\omega_x  \int dE \, \varrho_{x_0}\big(V_{x_0}(E,\{\omega_x\}_{x \neq x_0 }) \big)\,    (\cdots  )
\end{multline}
where in the last integral $V_{x_0}$ is regarded as a function of $E$ and 
$\{\omega_x\}_{x \neq x_0 }$, defined so that $E\in {\rm spec }  H_\mathcal{G,B}$, i.e.,  
\begin{equation} 
 V_{x_0}(E,\{\omega_x\}_{x \neq x_0 }) := - \langle \delta_{x_0} , (H^{(0)}_\mathcal{G,B} - E )^{-1} \delta_{x_0} \rangle^{-1} 
\end{equation}
where $H^{(0)}_\mathcal{G,B} $ is a precursor of $ H_\mathcal{G,B} $ 
with $ \omega_{x_0} = 0 $.   
 
The relation between  $H^{(0)}_\mathcal{G,B} $ and  $ H_\mathcal{G,B} $ 
is such that the Green function of the former appears as the eigenfunction of the latter 
when  $V_{x_0} $ is chosen as above, and in particular:
\begin{equation} 
  \frac{\psi_E(x)}{\psi_E(x_0)} = \frac{ \langle\delta_{x_0} , (H_\mathcal{G,B} - E )^{-1} \delta_{x} \rangle}{\langle \delta_{x_0} , (H_\mathcal{G,B} - E )^{-1} \delta_{x_0} \rangle } \, .
\end{equation} 
Furthermore, the ratio on the right does not depend on the value of $V_{x_0} $.  

For the correlation of eigenfunctions at energies in a Borel set $ I \subset \mathbb{R} $, 
which is the quantity of interest for us, the above considerations yield:  
\begin{equation}
  \mathbb{E}\Big[ \sum_{E \in {\rm spec}( H_\mathcal{G,B})\cap I }  \left| \psi_E(x_0) \right|  \; \left|\psi_E(x) \right| \Big] = \int_I  S(x_0,x;E) \, d E  \, .
\end{equation}
with 
\begin{equation}\label{def:efcorr}
  S(x_0,x;E) := \!\int \left|\frac{ \langle\delta_{x_0} , (H_\mathcal{G} - E )^{-1} \delta_{x} \rangle}{\langle \delta_{x_0} , (H_\mathcal{G} - E )^{-1} \delta_{x_0} \rangle} \right| 
  \, \varrho_{x_0}\big(V_{x_0}(E,\{\omega_y\}_{y \neq x_0 })\big) \, \prod_{y \neq x_0} \varrho_y (\omega_y) d\omega_y \, .
\end{equation}
This representation is obtained by averaging 
$\left| \frac{\psi_E(x)}{\psi_E(x_0)} \right|  \, \indfct_I(E) $ 
 with respect to the probability measure in \eqref{eqnasty}. 
    
In case of a backbone graph (in the sense of Definition~\ref{def:backbonegraph}) one may now further change variables in the integral in \eqref{def:efcorr} and introduce the 
Riccatti variables associated with $ x \in \mathcal{B} \setminus \{ x_0 \} $
\begin{equation}
 r_x :=  \left\{ \begin{array}{lr} 
   \frac{\psi_E(x-1)}{\psi_E(x)} \, , &  \quad x > x_0  \, , \\
    \frac{\psi_E(x+1)}{\psi_E(x)} \, , &  \quad x < x_0  \, .
    \end{array} \right.
\end{equation}
For  $
 H_\mathcal{G,B} = A + W + V $ these variables are related to the random variables $ \{\omega_x\}_{x\neq x_0} $ through the Schr\"odinger equation
$
  \omega_x = E - W_x - \frac{\psi_E(x+1)}{\psi_E(x)} - \frac{\psi_E(x-1)}{\psi_E(x)} $.
This implies that 
\begin{align}
&  S(x_0,x;E) = \int \left(| r_1 |^{-1} \, | r_2 |^{-1} \cdots | r_x |^{-1}\right) 
\varrho_{x_0}(E - W_{x_0} - r_{-1}^{-1}- r_{1}^{-1}) \notag \\
& \times \prod_{y > x_0 } \varrho_y (E - W_y - r_y - r_{y+1}^{-1}  ) \, d\omega_y \; \prod_{y < x_0 } \varrho_y (E - W_y - r_y - r_{y-1}^{-1}  )\, d\omega_y \, .
\end{align}

The above integral is identical to that in \cite[Eq.~(2.11)]{DeKuSo83}. It is proven there 
for the   case of interest to us, with  the random identically distributed,   
that there exists $ C < \infty $ such that for any $ E \in I $ 
\begin{equation}
 S(x_0,x;E) \leq  C  \, e^{- \lambda \, {\rm dist}(x_0, x)} \, .
\end{equation}
This completes the proof.
\end{proof} 

\subsection{Proof of Proposition~\ref{prop:DKS}}

Proposition~\ref{prop:DKS} which concerns the decay of the 
factional moments of the Green function will be deduced from 
Proposition~\ref{prop:DKS2} through rather general considerations relating that quantity to 
to the eigenfunction correlators.

\begin{lemma}
Let $ H $ be a self-adjoint operator with purely discrete spectrum with eigenvalues  $ E_1 \leq E_2 \leq \dots $ (counting multiplicity) and corresponding orthonormal set 
$\{ \psi_{E_n}\} $ of eigenfunctions. Then: 
\begin{enumerate}
\item For any pair of bounded intervals $ I, J \subset \mathbb{R} $, and $ s \in (0,1) $,  
\begin{equation}
  \int_I \left|\langle \delta_x , P_J(H) \, (H-E)^{-1}  \delta_y \rangle \right|^s d E \leq C_s(I) \, \left( \sum_{E_n \in J } | \psi_{E_n}(x)| \, | \psi_{E_n}(y)| \right)^s \, , 
\end{equation}
where $ C_s(I) := 4 s \int_\mathbb{R}  t^{s-1} \min\left\{ |I| , t^{-1} \right\} dt < \infty $. 
\item 
For any Borel $ J \subset \mathbb{R} $ and $ E \in J $ with $ {\rm dist}(E, \partial J ) > 0 $ one has 
\begin{align}
 &\left|\langle \delta_x , P_{J^c}(H) \, (H-E)^{-1}  \delta_y \rangle \right|^2 \notag \\
 & \qquad \leq \sum_{\substack{m \in \mathbb{Z} : \\ I_m \cap J = \emptyset}} \frac{1}{{\rm dist}(E,I_m)^2} \sum_{E_n  \in I_m } | \psi_{E_n}(x)| \, | \psi_{E_n}(y)|  
\end{align}
with $ I_m := (m,m+1] $.
\end{enumerate}
\end{lemma}
\begin{proof}
Abbreviating $ g(E) := \langle \delta_x , P_J(H) \, (H-E)^{-1}  \delta_y \rangle $, we start from the representation 
\begin{equation} \label{toomean} 
  \int_I \left|  g(E) \right|^s d E = s \int_0^\infty  t^{s-1} \, \left| \left\{ E \in I \, : \, | g(E) | > t \right\} \right| \, dt \, . 
\end{equation}
In terms of the spectral representation: $  g(E) = \sum_{E_n \in J } (E_n - E)^{-1} \psi_{E_n}(x) \, \overline{\psi_{E_n}(y)}   $.   After a decomposition of 
$ \psi_{E_n}(x) \, \overline{\psi_{E_n}(y)}  $ into  four terms corresponding to  positive, respectively non-positive, real and imaginary parts, one may apply  Boole's lemma \cite{Boo57}  to 
obtain: 
\begin{equation}  
  \left| \left\{ E \in I \, : \, | g(E) | > t \right\} \right| \leq 4 \, \,  \min\left\{ |I| ,  t^{-1} \sum_{E_n \in J } | \psi_{E_n}(x)| \, | \psi_{E_n}(y)| \right\} \, . 
\end{equation}
Substitution in \eqref{toomean}  yields the first assertion.

For a proof of the second bound we use the spectral decomposition and estimate  
\begin{align}
 & \left|\langle \delta_x , P_{J^c}(H) \, (H-E)^{-1}  \delta_y \rangle \right|  \leq \sum_{E_n \in J^c } \frac{| \psi_{E_n}(x)| \, | \psi_{E_n}(y)}{|E_n - E|} \notag \\
 & \leq \left(\sum_{E_n \in J^c } \frac{| \psi_{E_n}(x)| \, | \psi_{E_n}(y)}{|E_n - E|^2}\right)^{1/2}  \left(\sum_{E_n } | \psi_{E_n}(x)| \, | \psi_{E_n}(y)| \right)^{1/2} 
\end{align}
Using the Cauchy-Schwarz inequality and the fact that $ \sum_{E_n } | \psi_{E_n}(x)|^2 = 1 $, the last term is seen to be bounded by one.  
The claim then follows by decomposing the set $  J^c $ into a the union of sets $ I_m \cap J^c $ and estimating the denominator in the first factor on each of those sets.
\end{proof}

\begin{proof}[Proof of  Proposition~\ref{prop:DKS}] 
We pick $ J \supset I $ such that $ {\rm dist}(\partial J, \partial I ) > 0 $ and decompose the Green function into the two parts according to the previous lemma.  The estimates presented there yield the bound
\begin{align}\label{eq:G<EC}
& \int_I \mathbb{E}\left[ \left|\langle \delta_x , (H_{\mathcal{G},\mathcal{B}} -E)^{-1}  \delta_y \rangle \right|^s  \right] d E \notag \\ 
&\leq C \,  
  \sup_{m \in \mathbb{Z} }  \left(\mathbb{E}\left[ \sum_{E_n  \in (m,m+1] } | \psi_{E_n}(x)| \, | \psi_{E_n}(y)| \right]\right)^{s/2} 
\end{align}
with the constant $ C $ depending on $ I $ and $ s $. Thus 
Proposition~\ref{prop:DKS} can be deduced from Proposition~\ref{prop:DKS2}.
\end{proof}

It may be noted that  a converse relation also holds -- 
the expectation in the right-hand side of \eqref{eq:G<EC} can be bounded in terms of a suitably averaged fractional moment of the Green function \cite{A2}.

\section{Discussion}\label{sec:comments}

Our main result, Theorem~\ref{thm:main}, shows that regular tree 
graphs do not provide an example of the expected relation between the 
presence of ac spectrum for the infinite graph  and random-matrix 
like statistics in the spectra of the corresponding finite volume 
restrictions.
Let us therefore comment on a number of other directions in which it 
is natural to look for examples of such a  relation.  \\
\mbox{ }

As we saw, the negative result concerning the above relation
reflects the fact that a finite tree is mostly surface.  By
implication,  bulk averages of local quantities  yield results
representing the local mean not at sites deep within a tree but at
sites near the canopy.  In physicists discussions, the term `Bethe 
lattice average' is intended to reflect the average at sites deep 
within the tree, and  a
standard device is used for extracting it from the bulk sum.   For 
an extensive quantity such as $F_L = \Tr
F(H_{{\mathcal{H} }_L}) $ where
${\mathcal{H}} $ denotes the homogeneous tree
in which also the root has $K+1$ neighbors,
the `Bethe lattice average' $\langle F
\rangle_{BL}$ is obtained not by taking $\lim_{L \to \infty}
F_L / |{{\mathcal{H}}_L}| $, which gives the weighted canopy
average \eqref{eq:dos}, but rather (as in \cite{MilDer93}) through the limit:
\begin{equation} \label{Bethe}
\langle F \rangle_{BL} \ = \ \lim_{L \to \infty} \left( F_L - K \, 
F_{L-1} \right ) /2 \, .
\end{equation}
It would be of interest to see an adaptation of this approach
for some separation of the  statistics of eigenvalues  corresponding
to regions deep within the tree from the canopy average.
However, even for the mean density of states,  averaged over the 
disorder,  it remains to be
shown that the corresponding limit exists, and is given by a positive 
spectral measure.
Furthermore, it is not clear how to use an analog of \eqref{Bethe}
for specific realizations of an operator with disorder, as the latter
  ruins the homogeneity.   \\

Alternatively, one may look for graphs which have local tree 
structure without an obvious surface.
Let us briefly comment on results which relate to two such cases: the
random regular and the random Erd\H{o}s-R\'enyi graph also known as 
the sparse random matrix ensemble.

  The ensemble of random $ c$-regular graphs~\cite{Bol85}
consists of the
uniform probability measure on graphs on $ N \in \mathbb{N} $ vertices
where each vertex has $ c $ neighbors.
It is known that as $ N \to \infty $ the girth of the graph
(the minimal loop length) diverges in probability and
numerical simulations suggest \cite{JMRR99} that for large $ c $ the
eigenvalue spacing distribution of the adjacency operator approaches
that of the Gaussian Orthogonal Ensemble (GOE).

The ensemble of Erd\H{o}s-R\'enyi graphs results from the
complete graph on $  N \in \mathbb{N} $ vertices by removing bonds
with probability $ 1-p $.
     This ensemble is known to have a percolation transition with
an infinite tree-like connected component appearing as $ N \to \infty
$
     if the average connectivity $ c := p N $ is bigger than one.
The adjacency operator on these graphs is believed to exhibit a
quantum percolation transition, i.e.,
     the existence of extended eigenstates,
     at some value $ c > 1 $. Numerical \cite{Ev92,EE92,BG00} and
theoretical-physics calculations \cite{MiFy91} suggest that the
     eigenvalue spacing distribution of the adjacency operator
approaches GOE at least for large values of $ c $ (possibly depending
on $ N $).

Since the graphs in both ensembles do not show a an obvious
surface for finite $ N $, they may offer a natural setting for the
study of the relation between the extendedness of eigenstates of a
finite volume random Schr\"odinger operator and its level
statistics (a point which was also made  in private discussions
by T. Spencer).

\section*{Acknowledgement}
It is a pleasure to thank R. Sims and D. Jacobson for stimulating discussions
of related topics.  We also thank the referee for useful references
concerning the appearance of the canopy graph  in other studies.
Some of the work was done at the Weizmann
Institute (MA), at the Department of Physics of Complex Systems,
and at University of Erlangen-N\"urnberg (SW), Department of
Physics. We are grateful for the hospitality enjoyed there. This
work was supported in part by the NSF Grant DMS-0602360 and the
Deutsche Forschungsgemeinschaft (Wa~1699/1).


\bibliographystyle{plain}

\end{document}